# Tunneling escape of waves

David A. B. Miller[1], Zeyu Kuang[2], and Owen D. Miller[2]

[1]*Ginzton Laboratory, Stanford University, Stanford CA 94305, USA*

[2]*Department of Applied Physics and Energy Sciences Institute, Yale University, New Haven, Connecticut 06511, USA*

We solve a long-standing set of problems in optics and waves: why does a volume have only so many useful orthogonal wave channels in or out of it, why do coupling strengths fall off dramatically past this number, and, indeed, just what precisely defines that number? Increasingly in applications in communications, information processing, and sensing, in optics, acoustics, and electromagnetic waves generally, we need to understand this number. We can numerically find such channels for many problems, but more fundamentally, these questions have arguably never had a clear answer or physical explanation. We have found a simple and general result and intuition that lets us understand and bound this behavior for any volume. This is based on a tunneling that has been somewhat hidden in the mathematics of spherical waves: beyond a certain complexity of the wave, it must tunnel to escape the volume. By counting the number of waves that do not have to tunnel, we get a simple and precise number or bound for well-coupled channels, even for arbitrary volumes. The necessary tunneling for other waves explains the rapid fall-off in their coupling, and shows all such waves do escape to propagation to some degree after tunneling. This approach connects multipole expansions in electromagnetic antennas and nanophotonics smoothly to apparently evanescent waves in large optics. It works over all size scales, from nanophotonics, small radio-frequency antennas, or acoustic microphones and loudspeakers up to imaging optics with millions of channels, and gives a precise diffraction limit for any volume.

# Introduction

With emerging nanophotonics, increasingly we can design[1], [2] and fabricate sophisticated objects down to wavelength sizes or below. Growing bandwidth demands in radio-frequency (r.f.) wireless communications require we exploit spatial channels more effectively; resulting antenna systems are growing to increasingly sophisticated structures many wavelengths in size[3]–[5]. Information processing, such as for artificial intelligence, requires ever increasing numbers of channels that optics could provide, whether for improved inference in neural networks[6] or more generally in optical interconnects[7] and processing[8]. Numbers of strong channels can determine how much and what kind of information we can measure in sensing applications like microscopy or imaging or how many designable elements or basis functions we need in design. A core question, both for design and applications, is whether we can understand how many different (i.e., orthogonal) waves or channels can propagate in and out of objects or volumes. Such counting is increasingly relevant in optics now that we can control and detect light mode by mode[9]–[11]. This question has arguably never had a simple answer, especially as we move through objects of scales of a few wavelengths. Though we may believe that diffraction limits lie behind such counting, and we can certainly calculate the orthogonal channels and their coupling strengths for any specific case[12], we lack any simple general model and intuition for important key behaviors (see, e.g., examples in[12] and in many other analyses[13]–[18]): why do the coupling strengths tend to fall off rapidly past some number of well-coupled channels, and, indeed, just what defines that number?

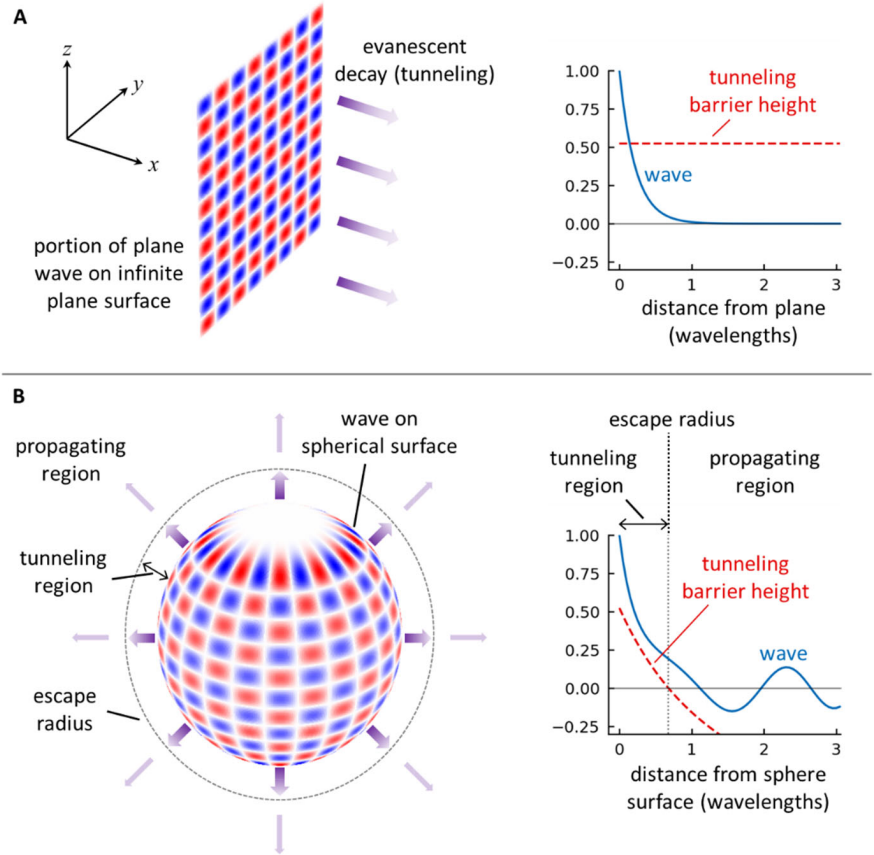

Fig. 1. Comparison of an evanescent plane wave and an initially tunneling spherical wave. (A) A plane wave emerging from an infinite surface in the $y - z$ plane, showing evanescent (tunneling) decay in the positive $x$ direction. (B) A spherical wave emerging from a spherical surface; initial tunneling behavior changes to propagating behavior past the escape radius. Both waves start with unit overall amplitude, at the plane surface for (A) and at the sphere surface for (B), and with the same decay in distance. The plane wave in (A) has a transverse pattern corresponding to $n_z = 1.034$ and $n_y = 0.674$ periods per wavelength in the $z$ and $y$ directions, respectively, equivalent to $k_z = 2\pi n_z$ and $k_y = 2\pi n_y$ radians per wavelength. The spherical wave in (B) has a spherical harmonic angular form with $n = 22$ and $m = 12$. The spherical surface has radius $r_o = 2.9$ wavelengths. The wave in (B) has the phase where it corresponds to the $C_n$ Riccati-Bessel function. The escape radius is $\sim 3.58$ wavelengths, so $\sim 0.68$ wavelengths larger than $r_o$. The tunneling barrier height starts at 0.524 in both cases, being $n_z^2 + n_y^2 - 1 \equiv [(k_z^2 + k_y^2)/k^2] - 1$ for plane waves and $[n(n+1)/(kr)^2] - 1$ for the spherical waves, for radial distance $r$.

We show here that there is a useful and unified approach with simple results and physical intuition. This approach spans continuously from sub-wavelength objects to large optical scales. A key concept is the idea of tunneling escape of spherical waves. We find the onset of this tunneling can be precisely defined, allowing a clear counting of strongly coupled channels that propagate without tunneling. Past this number, the tunneling behavior explains the fall-off and just how rapid it must be. The complementary problem of waves focusing into a volume similarly obeys a required onset of tunneling that explains the difficulty of focusing past diffraction limits. Since any finite volume can fit within some bounding sphere, we acquire simple upper bounds for coupling in and out of arbitrary volumes. The onset of tunneling also gives a precise definition of a diffraction limit for volumes.

## Spherical waves and tunneling

Just as plane waves describe waves from (infinite) plane surfaces, so spherical waves based on spherical harmonic and spherical Bessel functions usefully describe waves from spherical surfaces or finite objects. Their mathematics is well understood, e.g., as in scattering from spherical objects[19]–[23] and from multipole expansions of fields[24], [25]. (See supplementary text S1 and S2 for details.)

Fig. 1 illustrates the argument. For infinite plane waves (Fig. 1(a)), if the wave varies too rapidly in the $y$ and $z$ directions, we have evanescent decay in the $x$ direction. Such a truly evanescent (literally, "vanishing") wave never escapes to propagate in the $x$ direction. Fig. 1(b) shows a spherical wave that, at a radial distance $r_o$ from the center, has an amplitude in angle given by a spherical harmonic function of order $n$. Such spherical waves retain their angular shape as they expand radially.

Suppose this spherical wave initially varies too rapidly (in transverse directions, on the sphere surface) compared to a wavelength. Then, this wave still expands spherically, but must tunnel radially until it reaches some escape radius $r_{escn}$ for this $n$. By this point, its transverse variation is slow enough that it can start to propagate.

Though the wave amplitude remaining after tunneling may be quite small, that wave will continue expanding, settling to ~1/radius decay of its amplitude, as in a spherically expanding propagating wave. So, waves from finite bodies, even if they start out somewhat evanescently, do not remain so; at least to some degree, they escape to propagation.

### *Scalar waves*

To justify this argument, we analyze spherical waves, starting with the scalar case, with a Helmholtz wave equation

$$\nabla^2 U(\mathbf{r}) + k^2 U(\mathbf{r}) = 0 \qquad (1)$$

for a wave $U(\mathbf{r})$ of frequency $f_o$ in a uniform medium such as vacuum, air, or an isotropic dielectric. (For some wave velocity $v$, the wavevector magnitude $k = 2\pi/\lambda_o = \omega/v$, where the wavelength and angular frequency are $\lambda_o = v/f_o$ and $\omega = 2\pi f_o$, respectively.) Using complex waves with time-dependence $\exp(-i\omega t)$, a plane wave solution has the form $U(\mathbf{r}) \propto \exp(i\mathbf{k} \cdot \mathbf{r})$ in space, with wavevector $\mathbf{k} = k_x\hat{\mathbf{x}} + k_y\hat{\mathbf{y}} + k_z\hat{\mathbf{z}}$, where $k_x$, $k_y$, and $k_z$ are the wavevector components in the $x$, $y$, and $z$ directions (with corresponding unit vectors $\hat{\mathbf{x}}$, $\hat{\mathbf{y}}$, and $\hat{\mathbf{z}}$). Such a solution $\exp(i\mathbf{k} \cdot \mathbf{r}) \equiv \exp(ik_x x)\exp(ik_y y)\exp(ik_z z)$, of Eq. (1), implicitly in infinite space, is separable as $U(\mathbf{r}) = X(x)Y(y)Z(z)$, with $k_x^2 + k_y^2 + k_z^2 = k^2$. If $k_y^2 + k_z^2 > k^2$, then $X(x) \propto \exp(-\kappa x)$, where $\kappa = \sqrt{k_y^2 + k_z^2 - k^2}$; presuming any sources are on the "left", in the region with $x < 0$, we take the positive square root. This is the classic evanescent wave, decaying exponentially for all $x > 0$; note it never becomes a propagating wave for any positive $x$ and the change from propagating to evanescent is totally abrupt as soon as $k_y^2 + k_z^2 > k^2$.

We can write the separated equation for $X(x)$ (see supplementary text S1.1) as

$$-\frac{d^2\xi(\rho)}{d\rho^2} + V(\rho)\xi(\rho) = E\xi(\rho) \tag{2}$$

where $\rho = k_x x$ and $X(x) \equiv \xi(\rho)$, which is in the form of a one-dimensional Schrödinger equation (see, e.g., [26]), with a "potential" energy $V(\rho) = (k_y^2 + k_z^2)/k^2$ (here actually independent of $\rho$), and an "eigenenergy" $E = 1$. This rewriting as in Eq. (2) allows us to state the clear behavior that, if $V(\rho) < E$, we have propagating solutions, and if $V(\rho) > E$, we have tunneling solutions. (Other recent work[27] has explored an abstract quantum tunneling approach with finite plane surfaces, though our spherical approach here is different.)

The scalar Helmholtz equation (1) can also be solved in spherical polar coordinates, $r$, $\theta$ and $\phi$ (see, e.g., [25]) (see supplementary text S1.2). Specifically, one can look for separable solutions $U(\mathbf{r}) = R(r)Y(\theta,\phi)$. The angular solutions $Y(\theta,\phi)$ are the spherical harmonics $Y_{nm}(\theta,\phi)$, with $n = 0, 1, 2, \ldots$ and, using the complex forms, $-n \leq m \leq n$ (see, e.g.,[26]). The radial solutions $R(r)$ are the spherical Bessel functions $z_n(kr)$, also for $n = 0, 1, 2, \ldots$. Using a dimensionless radial variable $\rho = kr$, these radial functions satisfy the differential equation

$$\rho^2 \frac{d^2 z_n(\rho)}{d\rho^2} + 2\rho \frac{dz_n(\rho)}{d\rho} + \left(\rho^2 - n(n+1)\right)z_n(\rho) = 0 \tag{3}$$

Spherical Bessel functions of the first and second kinds, respectively $j_n(\rho)$ and $y_n(\rho)$, give two independent solutions of Eq. (3). For a given $n$, linear combinations of these are also solutions. In particular, the spherical Hankel function of the first kind

$$h_n^{(1)}(\rho) = j_n(\rho) + iy_n(\rho) \tag{4}$$

corresponds to outward propagating waves at large distances. (The notation $z_n(\rho)$ stands in for any of $j_n(\rho)$, $y_n(\rho)$, or $h_n^{(1)}(\rho)$ in Eq. (3).) All these solutions have an underlying $1/r$ or $1/\rho$ dependence at large $r$ or $\rho$, which corresponds to them falling off ultimately as spherically expanding waves. Indeed, specifically,

$$h_n^{(1)}(\rho) \to i\frac{\exp(i\rho)}{\rho} \equiv \frac{i}{k}\frac{\exp(ikr)}{r} \tag{5}$$

is a spherically expanding propagating wave as $r \to \infty$.

We can multiply by radius to take out this underlying $1/r$ or $1/\rho$ dependence, giving radial solutions then expressed using what are known as Riccati-Bessel functions $S_n(\rho) = \rho j_n(\rho)$ and $C_n(\rho) = -\rho y_n(\rho)$, and in particular, the "outgoing" Riccati-Bessel function

$$\xi_n(\rho) = \rho h_n^{(1)}(\rho) \equiv S_n(\rho) - iC_n(\rho) \tag{6}$$

Such functions can be convenient for viewing the wave more as a function of angle rather than transverse position at any radius.

There is, however, one other very important consequence of using a Riccati-Bessel functions like $\xi_n(\rho)$: they obey the Riccati-Bessel differential equation, which can be written

$$-\frac{d^2\xi_n(\rho)}{d\rho^2} + \frac{n(n+1)}{\rho^2}\xi_n(\rho) = \xi_n(\rho) \tag{7}$$

(Note that substituting from Eq. (6) in Eq. (7) will recover Eq. (3)). This equation is now in exactly the Schrödinger form as in Eq. (2). The "eigenenergy" $E = 1$ as before, but now the effective radial potential is

$$V(\rho) = \frac{n(n+1)}{\rho^2} \tag{8}$$

which falls off (as $1/\rho^2$) with radius $\rho$ or $r$ ($= \rho/k$). We can usefully now define the "escape radius"

$$\rho_{escn} = \sqrt{n(n+1)} \text{ or, equivalently, } r_{escn} = \frac{\sqrt{n(n+1)}}{k} \equiv \frac{\lambda_o}{2\pi}\sqrt{n(n+1)} \tag{9}$$

which marks the boundary between tunneling and propagating behavior for the wave, i.e., the point $r$ or $\rho$ at which $V(kr) = E$ $(=1)$. So, we have outgoing wave solutions

$$U_{nm}(\mathbf{r}) = h_n^{(1)}(kr)Y_{nm}(\theta,\phi) \equiv \frac{\xi_n(kr)}{kr}Y_{nm}(\theta,\phi) \quad n = 0,1,2..., \ -n \le m \le n \tag{10}$$

If an outward wave has an angular form $Y_{nm}(\theta,\phi)$, then for $r < r_{escn}$, this wave (in the Riccati-Bessel form $\xi_n(kr)$) is tunneling outwards, but once it passes the escape radius $r_{escn}$, it becomes propagating, escaping at least to some degree. This contrasts with (infinite) plane waves of the form $U(\mathbf{r}) \propto \exp(i\mathbf{k}\cdot\mathbf{r})$; if such waves start out as evanescent (so with $k_y^2 + k_z^2 > k^2$), they remain evanescent at all $x$, eventually vanishing completely.

The wave plotted in Fig. 1(b) is such a Riccati-Bessel function, showing the tunneling-like behavior up to the escape radius, and the propagating behavior for larger radii. (See fig. S5 in the supplementary text for the behavior as a function of time.)

Though this transition between tunneling and propagating behavior at the escape radius is clear from the differential equation (7), it is not at all obvious from the usual algebraic expressions for spherical Bessel or Riccati-Bessel functions, which involve series of inverse powers of the radius together with sine and cosine functions (see, e.g., [25], p. 426); this may be why this tunneling behavior is not already better known.

Incidentally, the scalar spherical waves found this way are also the waves associated with the communication modes[12], [13] between spherical surfaces (see supplementary text S3).

## *Vector electromagnetic waves*

There are three different forms of vector wave solutions to the vector Helmholtz equation (see supplementary text S2), with one "longitudinal" and two "transverse" polarizations. The angular aspects are describable based on three vector spherical harmonic functions. Importantly, the tunneling in the radial behavior $R(r)$ above for scalar waves persists into the vector cases.

Though the longitudinal wave exists for sound and elastic waves, electromagnetic waves have just two, transverse forms. Each of those is separable into radial and angular parts, with the radial parts obeying the same equation as the function $R(r)$ above, but with the angular part being a vector spherical harmonic function. Explicitly, for outgoing waves we have a set of "transverse electric" (TE) waves, with electric field (from Eq. (155) in the supplementary text)

$$\mathbf{E}_{nm}^{(TE)}(r,\theta,\phi) = i\sqrt{\frac{\mu}{\varepsilon}} h_n^{(1)}(kr)\mathbf{C}_{mn}(\theta,\phi) \equiv i\sqrt{\frac{\mu}{\varepsilon}} \frac{\xi_n(kr)}{kr}\mathbf{C}_{mn}(\theta,\phi) \quad n=1,2..., -n \leq m \leq n \quad (11)$$

and "transverse magnetic" (TM) waves with magnetic field (from Eq. (159) in the supplementary text)

$$\mathbf{H}_{nm}^{(TM)}(r,\theta,\phi) = i h_n^{(1)}(kr)\mathbf{C}_{mn}(\theta,\phi) \equiv i\frac{\xi_n(kr)}{kr}\mathbf{C}_{mn}(\theta,\phi) \quad n=1,2..., -n \leq m \leq n \quad (12)$$

Here, $\mathbf{C}_{mn}$ is the vector spherical harmonic function

$$\mathbf{C}_{mn}(\theta,\phi) = \nabla \times \left[\mathbf{r} Y_{nm}(\theta,\phi)\right] \equiv \nabla Y_{nm}(\theta,\phi) \times \mathbf{r} \quad n=1,2..., -n \leq m \leq n \quad (13)$$

Note explicitly that radial behavior is given by functions $h_n^{(1)}(kr)$ or $\xi_n(kr)$, just as in the scalar case. The main differences are that

(i) we have two waves for each choice of *n* and *m*, which we can view as being TE and TM polarized waves, respectively,

(ii) their angular form, which is vectorial, is based on the gradient $\nabla Y_{nm}$ of the spherical harmonic rather than just the (scalar) spherical harmonic $Y_{nm}$ directly. Since $Y_{nm}(\theta,\phi)$ only depends on angles, not radius, $\nabla Y_{nm}$ only has vector components in the $\hat{\theta}$ and $\hat{\phi}$ directions on the sphere surface, so it and the function $\mathbf{C}_{mn}(\theta,\phi)$ contain only transverse components – vectors that lie in the sphere surface,

(iii)  in contrast to the scalar solutions (10), there are electromagnetic waves for $n=0$.

The electric field $\mathbf{E}_{nm}^{(TM)}$ corresponding to $\mathbf{H}_{nm}^{(TM)}$ and the magnetic field $\mathbf{H}_{nm}^{(TE)}$ corresponding to $\mathbf{E}_{nm}^{(TE)}$ are given by Eqs. (158) and (154), respectively, in the supplementary text. The TE and TM waves for a given $n$ and $m$ describe perpendicularly polarized waves, with $\mathbf{E}_{nm}^{(TM)}$ perpendicular to $\mathbf{E}_{nm}^{(TE)}$ and similarly for the magnetic fields $\mathbf{H}_{nm}^{(TE)}$ and $\mathbf{H}_{nm}^{(TM)}$. The polarizations of both $\mathbf{E}_{nm}^{(TE)}$ and $\mathbf{H}_{nm}^{(TM)}$ are always transverse (so perpendicular to the radius vector $\mathbf{r}$), and the corresponding fields $\mathbf{E}_{nm}^{(TM)}$ and $\mathbf{H}_{nm}^{(TE)}$ are transverse in the far field, though they have radial components in the near field.

This approach corresponds exactly to the multipole expansion of outward-propagating electromagnetic fields. We write fields in forms equivalent to Jackson's definitions[25]; derivations of Eqs. (11) and (12) can be found there[25] (see supplementary text S2 for general vector wave derivations). Our tunneling analysis conceptually connects multipole and quasi-evanescent behaviors in one formalism. These vector spherical waves are the waves associated with the (vector) communication modes between spherical surfaces or volumes[14].

Note that our analysis also tells us that, to observe strong radiation from high $n$ multipoles from some object, to avoid having to tunnel to escape, it would have to be essentially of the scale of twice the escape radius (so the diameter of the bounding sphere) for that $n$.

## Counting waves from spherical surfaces

The escape radius allows a useful counting of waves associated with a spherical surface of radius $r_o$. There is a maximum value $n_p$ of $n$ for which all the associated spherical harmonic waves propagate without tunneling to escape, which requires $r_o > r_{escn}$ for a given $n$. From Eq. (9), solving the quadratic equation $n(n+1) = (kr_o)^2$, the largest $n$ for which $kr_o > \sqrt{n(n+1)}$ is

$$n_p(r_o) = \text{floor}\left[\sqrt{(1/4) + N_{SH}} - (1/2)\right] \qquad (14)$$

( $\text{floor}(u)$ is the largest integer $\leq u$.). Here $N_{SH}$ is the "spherical heuristic number"

$$N_{SH} = (kr_o)^2 \equiv \frac{4\pi r_o^2}{(\lambda_o^2/\pi)} = \frac{A_S}{(\lambda_o^2/\pi)} \qquad (15)$$

where $A_S = 4\pi r_o^2$ is the area of the spherical surface. (The concept of $N_{SH}$ was introduced empirically in [12]; this algebraic definition of $N_{SH}$ also emerged in analytic work on spherical waves[14].) Note that, for $kr_o < \sqrt{2}$, or, equivalently, $N_{SH} < 2$ or $r_o < r_{esc1}$, where

$$r_{esc1} = \lambda_o/(\sqrt{2}\,\pi) \simeq 0.225\lambda_o \qquad (16)$$

only the $n=0$ wave propagates without requiring tunneling escape. Since there are no $n=0$ electromagnetic waves, all electromagnetic waves from (bounding) volumes smaller than this radius $r_{esc1}$ must tunnel to escape (consistent with the well-known Chu antenna limit[28], which states that antenna Q-factor must increase for small antennas). Note again that there is no corresponding requirement for scalar waves, such as normal sound waves, which is consistent with small microphones and loudspeakers not having a corresponding limit and high Q-factor requirement.

We can now calculate the total number $N_{ps}$ of (scalar) spherical harmonic waves with values of $n$ up to (and including) $n_p$. Since there are $2n+1$ different $m$ values (and hence spherical harmonics $Y_{nm}$) for each $n$,

$$N_{ps} = 1+3+5+\cdots(2n_p+1) \equiv \sum_{q=0}^{n_p}(2q+1) = (n_p+1)^2 \tag{17}$$

For electromagnetic waves, with no $n=0$ waves, we remove that one wave, giving

$$N_p = N_{ps} - 1 = n_p(n_p+2) \tag{18}$$

which is the number of such electromagnetic waves per polarization.

Since the Riccati-Bessel $\xi_n(kr)$ functions take out the underlying $1/r$ decay of spherical waves, a simple metric for how effectively a spherical wave of index $n$ propagates outwards from a radius $r_o$ is the relative far-field magnitude squared

$$\gamma_n(kr_o) = |\xi_n(\infty)|^2 / |\xi_n(kr_o)|^2 \tag{19}$$

Incidentally, for a given $n$, this coupling $\gamma_n$ is the same for every $m$ value, which is surprising given the different associated shapes of $Y_{nm}$.

In Fig. 2 (a) and (c), we plot $\gamma_n(kr_o)$ against $n$, and in Fig. 2 (b) and (d) parametrically against the cumulative number $n(n+2)$ of electromagnetic waves per polarization (see Eq. (18)).(See also fig. S6 for curves like Fig. 2 (a) and (b) for different radii of spherical surfaces.) We see several interesting and useful behaviors in Fig. 2.

1) For small volumes – e.g., a few wavelengths or smaller in radius – relatively quite a few waves can escape by tunneling with usefully large (e.g., $>10^{-2}$) propagating amplitudes.
2) The spherical heuristic number $N_{SH}$ is a good approximation to the total number $N_p$ of waves (per polarization) that start out as propagating.
3) As the radius of the spherical volume increases, a smaller fraction of the waves can usefully escape by tunneling, compared to those that start out as propagating. So, the transition to waves that must tunnel to increasing weak escape is increasingly relatively abrupt for larger volumes. So, in practice, with increasing size, we tend towards a simpler categorization of waves being either propagating or (approximately) evanescent.

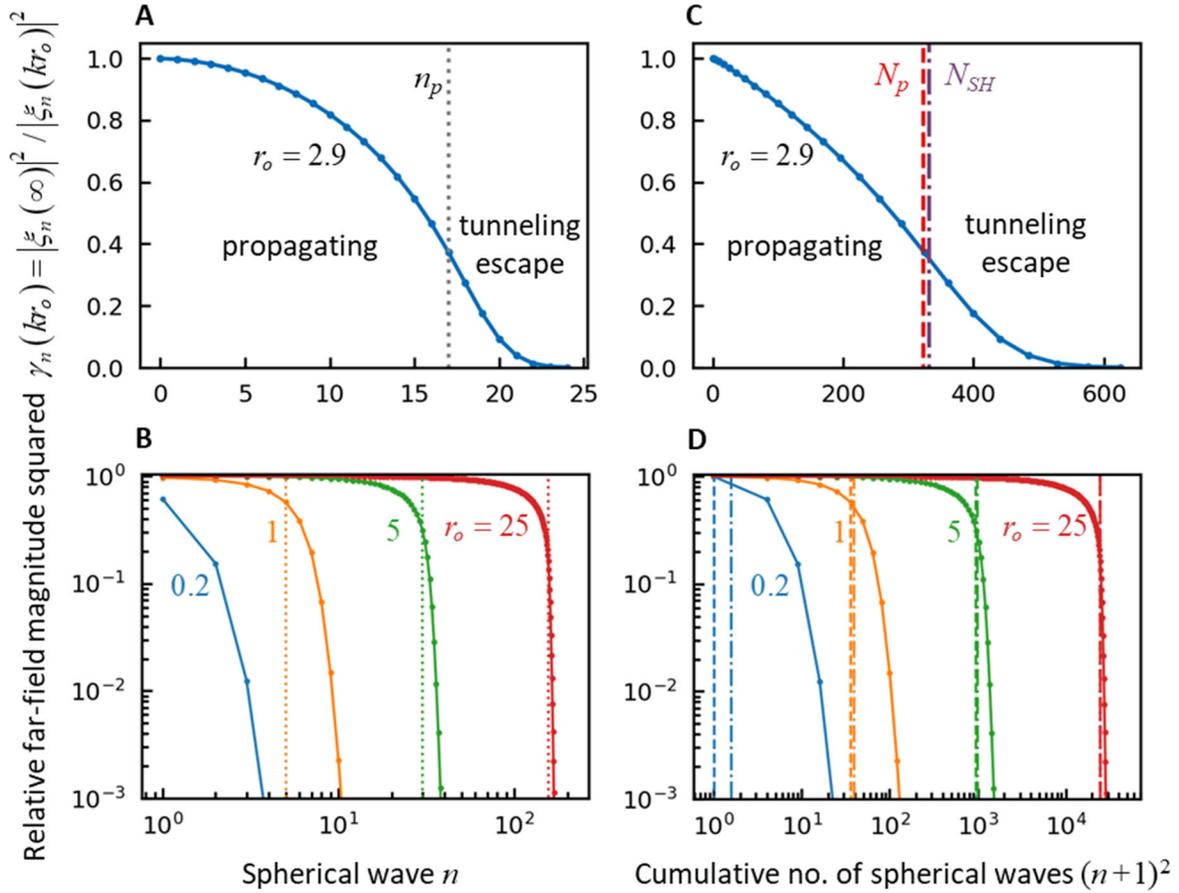

Fig. 2. Far-field spherical wave strength with increasing order $n$. Relative far-field strength $\gamma_n(kr_o) = |\xi_n(\infty)|^2 / |\xi_n(kr_o)|^2$ of the outgoing Riccati-Bessel function of order $n$ for various values of the starting radius $r_o$ in wavelengths, plotted against $n$ in (A) and (B), and, in (C) and (D), parametrically against the total number $n(n+2)$ of spherical waves for the electromagnetic case, per polarization, up to and including waves of order $n$. (Solid lines are to guide the eye.) $n_p$ (vertical dotted lines on (A) and (B)) is the largest $n$ for which all waves start out as propagating from the given radius $r_o$. $N_p = n_p(n_p+2)$ (vertical dashed lines on (C) and (D)) is the total number of such waves. $N_{SH} = (kr_o)^2$ (vertical dash-dot lines on (C) and (D)) are the spherical heuristic number. (A) and (C) use linear scales, with $r_o = 2.9$ wavelengths as in Fig. 1. (The $n = 0$ point, which does not exist for electromagnetic waves, is shown on (A) and (C).) For $r_o = 2.9$ wavelengths, $n_p = 17$; $N_p = 323$; $N_{SH} \simeq 332.01$. (B) and (D) show results for $r_o$ of 0.2, 1, 5, and 25 wavelengths on log scales. The values of $n_p$, $N_p$, and $N_{SH}$ for $r_o = 0.2, 1, 5, 25$ are, respectively, $n_p = 0, 5, 30, 156$; $N_p = 1, 35, 960, 24648$; $N_{SH} \simeq 1.58, 39.48, 986.96, 24674.01$.

The actual number of waves that propagate strictly without tunneling is a series of steps of integer height as a function of the radius of the spherical surface (Fig. 3). The number $N_{SH}$ is a continuous smooth function that passes through those steps (Fig. 3). Hence, it can overall be a good simple estimate for this number $N_p$.

Note that $N_{SH}$ in Eq. (15) is also the area $A_S = 4\pi r_o^2$ of the spherical surface divided by an area $\lambda_o^2 / \pi$. So $N_{SH}$ corresponds to one such wave for every $\lambda_o^2 / \pi$ of area on the sphere, connecting this spherical wave behavior to heuristic diffraction limits in conventional optical systems with more planar surfaces, which limit focal spots to ~ a (square) half wavelength because of diffraction. Note, too, that at moderate to large $n$, we have $n_p \simeq \sqrt{N_{SH}} = 2\pi r_o / \lambda_o$, which is the circumference of the sphere in wavelengths.

Fig. 3 also shows the number of waves that couple to the far field with coupling $\gamma_n$, Eq. (19), greater than specific values of 0.1 and 0.01. (These would correspond to horizontal lines at $10^{-1}$ and $10^{-2}$ on Fig. 2(d).) Some such waves will be tunneling to escape, but might still be practically useful, e.g., for communications or sensing. We see that there are quite significant numbers of such additional waves for small sphere radius, though relatively fewer for larger radii.

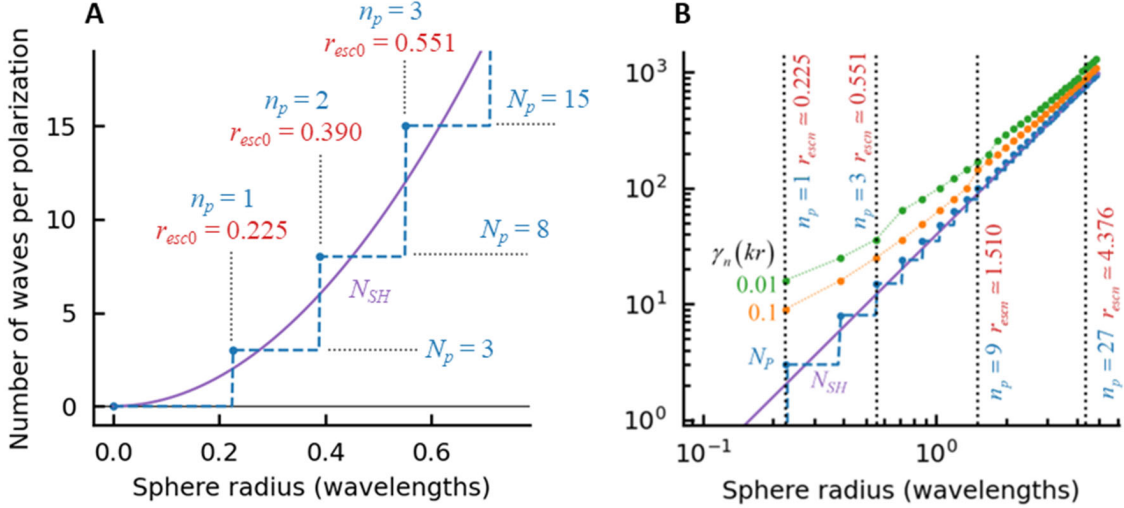

Fig. 3. Number of waves (per polarization for electromagnetic waves) that start out by propagating, as a function of the sphere radius. This number is the stepped dashed line. The solid line is the spherical heuristic number $N_{SH}$, showing it is a good approximation, even down to small radii. Also shown are corresponding values of $n_p$, $N_p$, and the escape radius $r_{escn}$ (in wavelengths). (A) Linear scale. (B) Log scale, with also example $\gamma_n(kr_o) = |\xi_n(\infty)|^2 / |\xi_n(kr_o)|^2$ values as the relative far-field magnitude squared of the spherical wave, with numbers of waves shown for $\gamma_n(kr_o) > 0.1$ and $> 0.01$ (dotted lines connecting points are just to guide the eye). Also indicated are several example $n_p$ values (1, 3, 9, and 27) and their corresponding escape radii $r_{escn}$, in wavelengths.

Though we consider only outgoing waves explicitly here, in practice these same numbers of waves also correspond essentially to the number of incoming waves that can penetrate into a spherical (bounding) volume (see supplementary text S4). Spherical waves beyond this limit will essentially reflect off an empty spherical volume of free space (becoming standing waves), so objects smaller than this volume are essentially invisible to, or "self-cloaked"[29] from, such waves.

## Waves from arbitrary volumes

The sets of scalar and electromagnetic spherical waves we have constructed are complete for describing any (outgoing) wave on a spherical surface. Hence, they can also describe any wave emerging from sources in a volume enclosed by that bounding spherical surface (Fig. 4). Since we have established how many orthogonal basis-set waves can emerge from this spherical surface without tunneling, we have established an upper bound on the maximum number of orthogonal waves that could emerge from the source volume without tunneling. We could also extend this estimate to allow for waves that could tunnel from the spherical surface to escape to some specified degree.

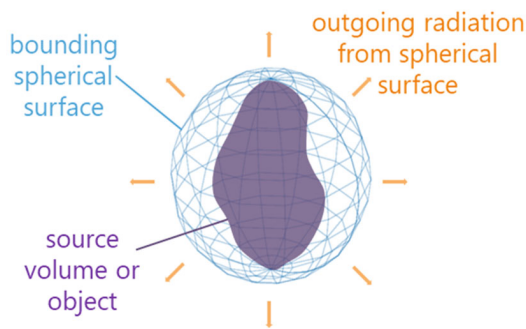

Fig. 4. A spherical bounding surface that just encloses some source volume or object. Any waves from the source volume that reach the spherical surface can be described in terms of spherical waves on that surface.

## Heuristic result for restricted solid angles

Once the spherical surface becomes several wavelengths or larger in size, the number of waves (per polarization) that propagate without tunneling becomes quite large. At radii of ~ 3, 5, and 10 wavelengths, for example, these numbers are ~ 350, 1000, and 4000, respectively. So, we can divide $N_{SH}$ by $4\pi$ steradians to obtain, for a spherical surface of radius $r_S$, the (approximate) number of propagating waves per unit solid angle.

$$N_\Omega \simeq \frac{\pi r_S^2}{\lambda_o^2} \qquad (20)$$

Imagine, then, some receiving surface of area $A_R$ at some (perpendicular) distance $L$ from the center of the spherical surface (Fig. 5), so subtending a solid angle $\Omega \simeq A_R / L^2$. Then from Eq. (20) we could estimate a total number of propagating wave channels

$$N_H = \frac{\pi r_S^2}{\lambda_o^2} \frac{A_R}{L^2} = \frac{A_T A_R}{\lambda_o^2 L^2} \tag{21}$$

where $A_T = \pi r_S^2$, which we note is the apparent circular cross-sectional area of the source volume (see Fig. 5). We note that Eq. (21) is exactly the number previously deduced as the "paraxial heuristic number" of well-coupled channels between planar source and receiver spaces[12] (Eq. (64)). (See also[12], Appendix A, for the many heuristic derivations of number.) So, if we count only those waves that do not require tunneling to escape, we can derive previous heuristic results for the "diffraction-limited" number of channels in paraxial optical systems. (Note, too, that $N_\Omega$, Eq. (20), is not itself restricted to paraxial cases.) These approaches could also be viewed as asking for the approximate number of these (spherical harmonic) basis functions required to adequately describe the resulting possible waves on the receiving surface from such a source volume.

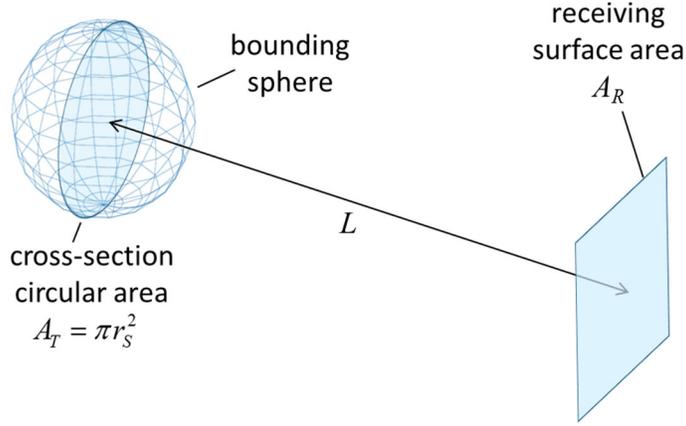

Fig. 5. Construction for waves to a finite receiving surface. A bounding spherical volume of radius $r_S$ surrounds a source of circular cross-sectional area $A_T = \pi r_S^2$ communicating to a receiving surface of area $A_R$ at a distance $L$.

## Discussion and conclusions

We have shown a unified way of thinking about waves in and out of volumes, from the propagating and evanescent fields of large optics to the multipole expansions of antennas and nanophotonics. This is based rigorously on the spherical waves associated with the spherical bounding surface round some volume or object. For all volumes from approximately a wavelength scale and upwards, a maximum number of well-coupled waves or orthogonal channels is understood as those that do not have to tunnel to escape the spherical surface. This onset of tunneling corresponds to a "knee" in coupling strength after which coupling falls rapidly because of the tunneling. A corresponding escape radius $r_{escn} = \left(\sqrt{n(n+1)}\right)\lambda_o / 2\pi$ characterizes the largest order $n$ of spherical wave of wavelength $\lambda_o$ that can propagate from such a spherical surface without initially tunneling. A spherical heuristic number $N_{SH}$, corresponding to one wave for every $\lambda_o^2/\pi$ of area on the spherical surface, approximately but usefully characterizes the number of well-coupled waves (per polarization) and the position of the "knee" in coupling strength. With increasing radius of the volume, the relative fall-off from tunneling becomes progressively more abrupt, asymptoting towards the complete abruptness of the onset of truly evanescent behavior for infinite plane waves. Note, though, that such truly evanescent waves are an artifact of the assumption of infinite plane waves; all corresponding waves from finite bodies eventually escape to some degree by tunneling. (Note, incidentally, that similar radial tunneling, propagating, and escape behavior can be derived for circular and cylindrical waves; see supplementary text S5.)

Finally, this approach allows us to propose a precise definition of the diffraction limit: For a wave interacting with a volume, the wave passes the diffraction limit if any spherical component of the wave must tunnel to enter or leave the bounding spherical surface enclosing the volume.

# References

[1] Y. Jiao, S. Fan, and D. A. B. Miller, "Demonstration of systematic photonic crystal device design and optimization by low-rank adjustments: an extremely compact mode separator," *Opt. Lett.*, vol. 30, no. 2, p. 141, Jan. 2005, doi: 10.1364/OL.30.000141.

[2] S. Molesky, Z. Lin, A. Y. Piggott, W. Jin, J. Vucković, and A. W. Rodriguez, "Inverse design in nanophotonics," *Nature Photonics*, vol. 12, no. 11, pp. 659–670, 2018, doi: 10.1038/s41566-018-0246-9.

[3] C. Ouyang, Y. Liu, X. Zhang, and L. Hanzo, "Near-Field Communications: A Degree-of-Freedom Perspective." arXiv, Aug. 02, 2023. doi: 10.48550/arXiv.2308.00362.

[4] T. Gong *et al.*, "Holographic MIMO Communications: Theoretical Foundations, Enabling Technologies, and Future Directions." TechRxiv, Apr. 29, 2023. doi: 10.36227/techrxiv.21669656.v2.

[5] Z. Wang *et al.*, "Extremely Large-Scale MIMO: Fundamentals, Challenges, Solutions, and Future Directions," *IEEE Wireless Communications*, pp. 1–9, 2023, doi: 10.1109/MWC.132.2200443.

[6] G. Wetzstein *et al.*, "Inference in artificial intelligence with deep optics and photonics," *Nature*, vol. 588, no. 7836, Art. no. 7836, Dec. 2020, doi: 10.1038/s41586-020-2973-6.

[7] D. A. B. Miller, "Attojoule Optoelectronics for Low-Energy Information Processing and Communications," *J. Lightwave Technol.*, vol. 35, no. 3, pp. 346–396, Feb. 2017, doi: 10.1109/JLT.2017.2647779.

[8] P. L. McMahon, "The physics of optical computing," *Nat Rev Phys*, Oct. 2023, doi: 10.1038/s42254-023-00645-5.

[9] D. A. B. Miller, "Self-configuring universal linear optical component," *Photon. Res.*, vol. 1, no. 1, pp. 1–15, Jun. 2013, doi: 10.1364/PRJ.1.000001.

[10] N. K. Fontaine *et al.*, "Photonic Lanterns, 3-D Waveguides, Multiplane Light Conversion, and Other Components That Enable Space-Division Multiplexing," *Proceedings of the IEEE*, pp. 1–14, 2022, doi: 10.1109/JPROC.2022.3207046.

[11] W. Bogaerts *et al.*, "Programmable photonic circuits," *Nature*, vol. 586, no. 7828, Art. no. 7828, Oct. 2020, doi: 10.1038/s41586-020-2764-0.

[12] D. A. B. Miller, "Waves, modes, communications, and optics: a tutorial," *Adv. Opt. Photon.*, vol. 11, no. 3, pp. 679–825, Sep. 2019, doi: 10.1364/AOP.11.000679.

[13] D. A. B. Miller, "Communicating with waves between volumes: evaluating orthogonal spatial channels and limits on coupling strengths," *Appl. Opt.*, vol. 39, no. 11, p. 1681, Apr. 2000, doi: 10.1364/AO.39.001681.

[14] Z. Kuang, D. A. B. Miller, and O. D. Miller, "Bounds on the Coupling Strengths of Communication Channels and Their Information Capacities." arXiv, May 10, 2022. doi: 10.48550/arXiv.2205.05150.

[15] E. A. Sekehravani, G. Leone, and R. Pierri, "Evaluation of the Number of Degrees of Freedom of the Field Scattered by a 3D Geometry," *Sensors*, vol. 23, no. 8, Art. no. 8, Jan. 2023, doi: 10.3390/s23084056.

[16] A. Pizzo and A. Lozano, "On Landau's Eigenvalue Theorem for Line-of-Sight MIMO Channels," *IEEE Wireless Communications Letters*, pp. 1–1, 2022, doi: 10.1109/LWC.2022.3208322.

[17] R. Solimene, M. A. Maisto, G. Romeo, and R. Pierri, "On the Singular Spectrum of the Radiation Operator for Multiple and Extended Observation Domains," *International Journal of Antennas and Propagation*, vol. 2013, p. e585238, Jul. 2013, doi: 10.1155/2013/585238.

[18] M. D. Migliore, "On the role of the number of degrees of freedom of the field in MIMO channels," *IEEE Transactions on Antennas and Propagation*, vol. 54, no. 2, pp. 620–628, Feb. 2006, doi: 10.1109/TAP.2005.863108.

[19] G. Mie, "Beiträge zur Optik trüber Medien, speziell kolloidaler Metallösungen," *Annalen der Physik*, vol. 330, no. 3, pp. 377–445, 1908, doi: 10.1002/andp.19083300302.

[20] P. Debye, "Der Lichtdruck auf Kugeln von beliebigem Material," *Annalen der Physik*, vol. 335, no. 11, pp. 57–136, 1909, doi: 10.1002/andp.19093351103.

[21] I. M. Hancu, A. G. Curto, M. Castro-López, M. Kuttge, and N. F. van Hulst, "Multipolar Interference for Directed Light Emission," *Nano Lett.*, vol. 14, no. 1, pp. 166–171, Jan. 2014, doi: 10.1021/nl403681g.

[22] D. Tzarouchis and A. Sihvola, "Light Scattering by a Dielectric Sphere: Perspectives on the Mie Resonances," *Applied Sciences*, vol. 8, no. 2, Art. no. 2, Feb. 2018, doi: 10.3390/app8020184.

[23] A. Dorodnyy, J. Smajic, and J. Leuthold, "Mie Scattering for Photonic Devices," *Laser & Photonics Reviews*, p. 2300055, 2023, doi: 10.1002/lpor.202300055.

[24] W. W. Hansen, "A New Type of Expansion in Radiation Problems," *Phys. Rev.*, vol. 47, no. 2, pp. 139–143, Jan. 1935, doi: 10.1103/PhysRev.47.139.

[25] J. D. Jackson, *Classical Electrodynamics*, Third Edition. John Wiley & Sons, Inc., 1999.

[26] D. A. B. Miller, *Quantum Mechanics for Scientists and Engineers*. Cambridge University Press, 2008.

[27] S. C. Creagh and G. Gradoni, "Slepian eigenvalues as tunnelling rates," *Annals of Physics*, vol. 449, p. 169204, Feb. 2023, doi: 10.1016/j.aop.2022.169204.

[28] H. Wheeler, "Small antennas," *IEEE Transactions on Antennas and Propagation*, vol. 23, no. 4, pp. 462–469, Jul. 1975, doi: 10.1109/TAP.1975.1141115.

[29] D. A. B. Miller, "On perfect cloaking," *Opt. Express*, vol. 14, no. 25, p. 12457, Dec. 2006, doi: 10.1364/OE.14.012457.

[30] J. A. Stratton, *Electromagnetic Theory*. McGraw-Hill, 1941.

[31] P. M. Morse and H. Feshbach, *Methods of Theoretical Physics Vol. 1*. McGraw-Hill, 1953.

[32] C. F. Bohren and D. R. Huffman, *Absorption and Scattering of Light by Small Particles*. Wiley, 1983.

[33] L. Tsang, J. A. Kong, and K.-H. Ding, *Scattering of Electromagnetic Waves: Theories and Applications*. New York, USA: John Wiley & Sons, Inc., 2000. doi: 10.1002/0471224286.

[34] K. Baek, Y. Kim, S. Mohd-Noor, and J. K. Hyun, "Mie Resonant Structural Colors," *ACS Appl. Mater. Interfaces*, vol. 12, no. 5, pp. 5300–5318, Feb. 2020, doi: 10.1021/acsami.9b16683.

[35] P. M. Morse and H. Feshbach, *Methods of Theoretical Physics Vol. 2*. McGraw-Hill, 1953.

[36] C. J. Bouwkamp and H. B. G. Casimir, "On multipole expansions in the theory of electromagnetic radiation," *Physica*, vol. 20, no. 1, pp. 539–554, Jan. 1954, doi: 10.1016/S0031-8914(54)80068-1.

[37] G. B. Arfken, H. J. Weber, and F. E. Harris, *Mathematical Methods for Physicists*, 7th Edition. Elsevier, 2013.

[38] G. W. Hanson and A. B. Yakovlev, *Operator Theory for Electromagnetics*. New York, NY: Springer New York, 2002. doi: 10.1007/978-1-4757-3679-3.

[39] D. A. Varshalovich, A. N. Moskalev, and V. K. Khersonskii, *Quantum Theory of Angular Momentum*. World Scientific, 1988. doi: 10.1142/0270.

# Acknowledgements

This work was supported by AFOSR grant FA9550-21-1-0312. References [30]–[39]

# Supplementary text

Waves in spherical coordinates are fundamentally well understood, though these details may be less commonly known. In this paper, we need to rely on some quite detailed aspects. We also need to deduce some basic and less common results with some rigor, especially as we consider spherical vector waves. We start with an extended discussion of scalar waves in spherical coordinates (Section S1). In Section S2, we give the derivation of the full behavior of vector wave solutions of the vector Helmholtz wave equation, including the special case of electromagnetic waves. This material can also be found in various forms in texts, with Refs.[25], [30]–[33] giving treatments that include the discussion of both scalar and vector cases. In Section S3, we show that the scalar wave solutions we have derived correspond to the communication modes[12], [13] between spherical shells. (Other work by us has shown the vector wave solutions similarly correspond to such communication modes in that case[14].) In section S4, we discuss how our approach applies to inward waves also. Section S5 shows that similar phenomena, with escape radius and radial tunneling, are found also in the radial behavior of cylindrical waves and the corresponding two-dimensional (planar) circular waves.

## S1 Scalar waves in spherical coordinates

We are interested here in scalar and vector waves in a uniform, lossless, isotropic medium. Vector waves in spherical coordinates are more complicated, especially in their general forms. Fortunately, however, the scalar solutions also form the basis for discussing the vector case, so we start with a simple scalar wave equation of the form

$$\nabla^2 W(\mathbf{r},t) - \frac{1}{v^2}\frac{\partial^2 W(\mathbf{r},t)}{\partial t^2} = 0 \qquad (22)$$

where $\mathbf{r}$ is the (vector) position in space relative to some origin, $t$ is time, $W(\mathbf{r},t)$ is the wave amplitude, and $v$ is the wave velocity. For electromagnetic waves, we would have $1/v^2 = \varepsilon\mu$ where $\varepsilon$ and $\mu$ are, respectively, the permittivity and permeability of the medium.

We consider waves of one frequency, $f_o$. The wavelength for such a wave would be $\lambda = v/f_o$. For algebraic convenience, we work with the angular frequency $\omega = 2\pi f_o$ and the wavevector magnitude $k = 2\pi/\lambda = \omega/c$, and we choose to write the corresponding time-dependence in the complex form $\exp(-i\omega t)$, which leads to

$$W(\mathbf{r},t) = \exp(-i\omega t)U(\mathbf{r}) \qquad (23)$$

with the understanding that we can take the real part at the end if we want to describe normal classical waves. The choice of the minus sign in the exponent in $\exp(-i\omega t)$ leads to the result that a wave of spatial form $\exp(ikr)/r$ will represent an outgoing spherical wave, for example.

$U(\mathbf{r})$ now represents the amplitude in space of this oscillation at frequency $f_o$ (or, equivalently at angular frequency $\omega$). Substituting Eq. (23) in Eq. (22) gives the Helmholtz wave equation as in the main text (Eq. (1))

$$\nabla^2 U(\mathbf{r}) + k^2 U(\mathbf{r}) = 0 \qquad (24)$$

## S1.1 Separation of the solutions - plane waves

For reference, and for a simple introduction or reminder in separating solutions, we start by showing the separation of solutions in Cartesian $(x, y, z)$ coordinates, which corresponds to plane wave solutions. With

$$\nabla^2 \equiv \frac{\partial^2}{\partial x^2} + \frac{\partial^2}{\partial y^2} + \frac{\partial^2}{\partial z^2} \tag{25}$$

we can formally propose a solution of the form

$$U(\mathbf{r}) = X(x)Y(y)Z(z) \tag{26}$$

Then

$$Y(y)Z(z)\frac{\partial^2 X(x)}{\partial x^2} + X(x)Z(z)\frac{\partial^2 Y(y)}{\partial y^2} + X(x)Y(y)\frac{\partial^2 Z(z)}{\partial z^2} = -k^2 X(x)Y(y)Z(z) \tag{27}$$

Dividing by $-X(x)Y(y)Z(z)$ and rearranging gives

$$-\frac{1}{X(x)}\frac{\partial^2 X(x)}{\partial x^2} = k^2 + \frac{1}{Y(y)}\frac{\partial^2 Y(y)}{\partial y^2} + \frac{1}{Z(z)}\frac{\partial^2 Z(z)}{\partial z^2} \tag{28}$$

As usual in such separation arguments, we note that nothing on the left of the equation depends on $y$ or $z$, and nothing on the right depends on $x$. Therefore, both sides must equal some ("separation") constant, which here, without loss of generality, we can choose to write as $k_x^2$. (We have not yet assigned any meaning to this way of writing the constant – it is just some number at this stage.) Hence, using the left side of Eq. (28) with this separation constant, and multiplying by $X(x)$ on both sides, we have

$$-\frac{d^2 X(x)}{dx^2} = k_x^2 X(x) \tag{29}$$

where we note we can use ordinary rather than partial derivatives because we now only have one variable in the equation. Solutions of this equation can be written in the form

$$X(x) = \exp(ik_x x) \tag{30}$$

where $k_x$ can take positive or negative values (and/or imaginary values), and arbitrary linear combinations of these are also solutions. Similarly, for $y$ and $z$ we have, with separation constants $k_y^2$ and $k_z^2$

$$Y(y) = \exp(ik_y y) \tag{31}$$

$$Z(z) = \exp(ik_z z) \tag{32}$$

Substituting using Eq. (29) and the corresponding similar equations for $Y(y)$ and $Z(z)$ back into Eq. (27) gives the necessary relation

$$k_x^2 + k_y^2 + k_z^2 = k^2 \tag{33}$$

We can also choose to write the overall separated solution as in Eq. (26) as

$$U(\mathbf{r}) = \exp(ik_x x)\exp(ik_y y)\exp(ik_z z) \equiv \exp(i\mathbf{k}\cdot\mathbf{r}) \tag{34}$$

with $\mathbf{k} = k_x\hat{\mathbf{x}} + k_y\hat{\mathbf{y}} + k_z\hat{\mathbf{z}}$ ($\hat{\mathbf{x}}$, $\hat{\mathbf{y}}$, and $\hat{\mathbf{z}}$ being the unit vectors in the $x$, $y$, and $z$ directions, respectively), which are the usual plane wave solutions. Note that, if $k_y^2 + k_z^2 > k^2$, then the solution in the $x$ direction is $X(x) \propto \exp(-\kappa x)$, where $\kappa = \sqrt{k_y^2 + k_z^2 - k^2}$; we take the positive square root (and hence a decaying exponential) on the presumption that any sources generating this wave are all on the "left", that is for $x < 0$. We describe such a wave as decaying evanescently (or exponentially) in the $x$ direction.

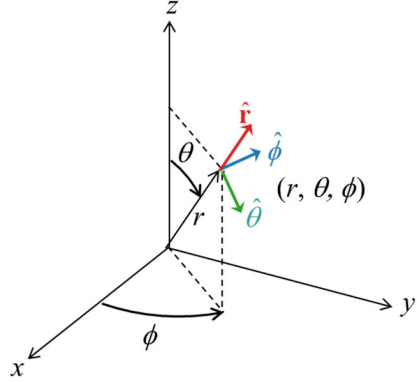

Fig. S1. Spherical polar coordinates $(r, \theta, \phi)$ for a point in space, relative to conventional $x$, $y$, and $z$ coordinate directions. Unit vectors $\hat{\mathbf{r}}$, $\hat{\theta}$, and $\hat{\phi}$ are also shown.

Formally, then, we can rewrite Eq. (29) as

$$-\frac{d^2 X(x)}{dx^2} = \left[k^2 - \left(k_y^2 + k_z^2\right)\right] X(x) \tag{35}$$

So

$$-\frac{d^2 X(x)}{dx^2} + \left(k_y^2 + k_z^2\right) X(x) = k^2 X(x) \tag{36}$$

Writing $\rho = kx$, $V(\rho) = (k_y^2 + k_z^2)/k^2$ (which is independent of $\rho$ in this case), and $E = 1$, we obtain the Schrödinger equation form of Eq. (2) in the main text.

## S1.2 Separation of the solutions - spherical waves

Having understood separation of solutions in Cartesian coordinates, now we consider spherical polar coordinates $r$, $\theta$, and $\phi$, as illustrated in Fig. S1. In such coordinates

$$\nabla^2 U \equiv \frac{1}{r^2} \frac{\partial}{\partial r}\left(r^2 \frac{\partial U}{\partial r}\right) + \frac{1}{r^2 \sin\theta} \frac{\partial}{\partial \theta}\left(\sin\theta \frac{\partial U}{\partial \theta}\right) + \frac{1}{r^2 \sin^2\theta} \frac{\partial^2 U}{\partial \phi^2} \tag{37}$$

For separation in spherical coordinates, we propose a form

$$U(\mathbf{r}) = R(r)\Theta(\theta)\Phi(\phi) \tag{38}$$

We substitute this form, together with the expression (37) for $\nabla^2 U$, into the wave equation (22), obtaining

$$\frac{\Theta(\theta)\Phi(\phi)}{r^2} \frac{\partial}{\partial r}\left(r^2 \frac{\partial R(r)}{\partial r}\right) + \frac{R(r)\Phi(\phi)}{r^2 \sin\theta} \frac{\partial}{\partial \theta}\left(\sin\theta \frac{\partial \Theta(\theta)}{\partial \theta}\right) + \frac{R(r)\Theta(\theta)}{r^2 \sin^2\theta} \frac{\partial^2 \Phi(\phi)}{\partial \phi^2} \\ + k^2 R(r)\Theta(\theta)\Phi(\phi) = 0 \tag{39}$$

Dividing by $U(\mathbf{r}) = R(r)\Theta(\theta)\Phi(\phi)$, multiplying by $\sin^2\theta$ and moving the second term to the right of the equation gives

$$\frac{\sin^2\theta}{R(r)} \frac{\partial}{\partial r}\left(r^2 \frac{\partial R(r)}{\partial r}\right) + \frac{\sin\theta}{\Theta(\theta)} \frac{\partial}{\partial \theta}\left(\sin\theta \frac{\partial \Theta(\theta)}{\partial \theta}\right) + k^2 r^2 \sin^2\theta = \frac{-1}{\Phi(\phi)} \frac{\partial^2 \Phi(\phi)}{\partial \phi^2} = m^2 \tag{40}$$

Here we have also taken the additional step, as usual in separation of variables, that, since nothing on the left of the equation depends on $\phi$, and nothing on the right of the equation depends on $r$ or $\theta$, both must equal some (separation) constant, which for future convenience we write as $m^2$. (Though $m$ will later be shown to be integer in a specific range, for the moment we have not restricted its value in any way.) This therefore has given us our first separated equation

$$\frac{d^2 \Phi(\phi)}{d\phi^2} = -m^2 \Phi(\phi) \tag{41}$$

which can now be written as a simple differential equation rather than a partial differential equation because there is only one variable, here $\phi$.

Now, taking the left hand side of Eq. (40), dividing by $\sin^2\theta$ and rearranging gives

$$\frac{1}{R(r)} \frac{\partial}{\partial r}\left(r^2 \frac{\partial R(r)}{\partial r}\right) + k^2 r^2 = \frac{m^2}{\sin^2\theta} - \frac{1}{\Theta(\theta)\sin\theta} \frac{\partial}{\partial \theta}\left(\sin\theta \frac{\partial \Theta(\theta)}{\partial \theta}\right) = n(n+1) \tag{42}$$

where again, now because nothing on the left depends on $\theta$ and nothing on the right depends on $r$, both must equal a constant, which for future convenience we choose to write as $n(n+1)$. (Again, later we will restrict $n$ to being an integer in a specific range, but so far there is no restriction on it.) So, we now have a two additional separated equations

$$\frac{1}{\sin\theta} \frac{d}{d\theta}\left(\sin\theta \frac{d\Theta(\theta)}{d\theta}\right) + \left(n(n+1) - \frac{m^2}{\sin^2\theta}\right)\Theta(\theta) = 0 \tag{43}$$

which we will see has solutions based on associated Legendre functions, and

$$\frac{d}{dr}\left(r^2 \frac{dR(r)}{dr}\right) + \left(k^2 r^2 - n(n+1)\right) R(r) = 0 \tag{44}$$

whose solutions we will see are spherical Bessel functions. Eq. (44) is more commonly written by performing the first derivative in the expression, and changing to a function in the form

$$z_n(\rho) \equiv z_n(kr) \equiv R(r) \tag{45}$$

with $\rho \equiv kr$, to give the equivalent equation

$$\rho^2 \frac{d^2 z_n(\rho)}{d\rho^2} + 2\rho \frac{dz_n(\rho)}{d\rho} + \left(\rho^2 - n(n+1)\right) z_n(\rho) = 0 \tag{46}$$

Equations (41), (44), and (44) (or (46)) are now the separated equations for the "$\phi$", "$\theta$", and "$r$" variables, respectively. We can now proceed to find consistent solutions for all three equations at once.

*Solving the $\theta$ equation*

By changing variables to $\eta = \cos\theta$ in Eq.(43), so changing to some function

$$P_n^m(\eta) \equiv P_n^m(\cos\theta) \equiv \Theta(\theta)$$

and noting that $d\cos\theta/d\theta = -\sin\theta$, Eq. (43) becomes

$$\frac{d}{d\eta}\left((1-\eta^2)\frac{dP_n^m(\eta)}{d\eta}\right) + \left(n(n+1) - \frac{m^2}{(1-\eta^2)}\right) P_n^m(\eta) = 0 \tag{47}$$

By performing the first differentiation, Eq. (47) is often written in the equivalent form

$$(1-\eta^2)\frac{d^2}{d\eta^2} P_n^m(\eta) - 2\eta \frac{dP_n^m(\eta)}{d\eta} + \left(n(n+1) - \frac{m^2}{(1-\eta^2)}\right) P_n^m(\eta) = 0 \tag{48}$$

Eq. (47) (or Eq. (48)) has meaningful solutions if we choose $n$ as zero or a positive integer, and if $|m|$ is an integer less than or equal to $n$, i.e.,

$$n = 0, 1, 2, \ldots \tag{49}$$

$$m = -n, \cdots, 0, \ldots n \tag{50}$$

The resulting solutions $P_n^m(\eta)$ are the associated Legendre functions. (These are sometimes called the associated Legendre polynomials, but they are not (finite length) polynomials if $m$ is odd.)

One way of defining $P_n^m(\eta)$ (see, e.g.,[25] p. 108) is through an extension of the Rodrigues formula

$$P_n^m(\eta) = \frac{(-1)^m}{2^n n!} (1-\eta^2)^{m/2} \frac{d^{n+m}}{d\eta^{n+m}} \left[(\eta^2 - 1)^n\right] \tag{51}$$

We can also consider these solutions with only 0 or positive $m$ because, other than for a prefactor, these functions are the same for positive and negative $m$. Explicitly,

$$P_n^{-m}(\eta) = (-1)^m \frac{(n-m)!}{(n+m)!} P_n^m(\eta) \tag{52}$$

In this form, the orthogonality (and normalization) integral for these functions over a range $-1 \leq \eta \leq 1$ (which is the only range of interest to us) is

$$\int_{-1}^{1} P_q^m(\eta) P_n^m(\eta) d\eta = \frac{2}{2n+1} \frac{(n+m)!}{(n-m)!} \delta_{qn} \tag{53}$$

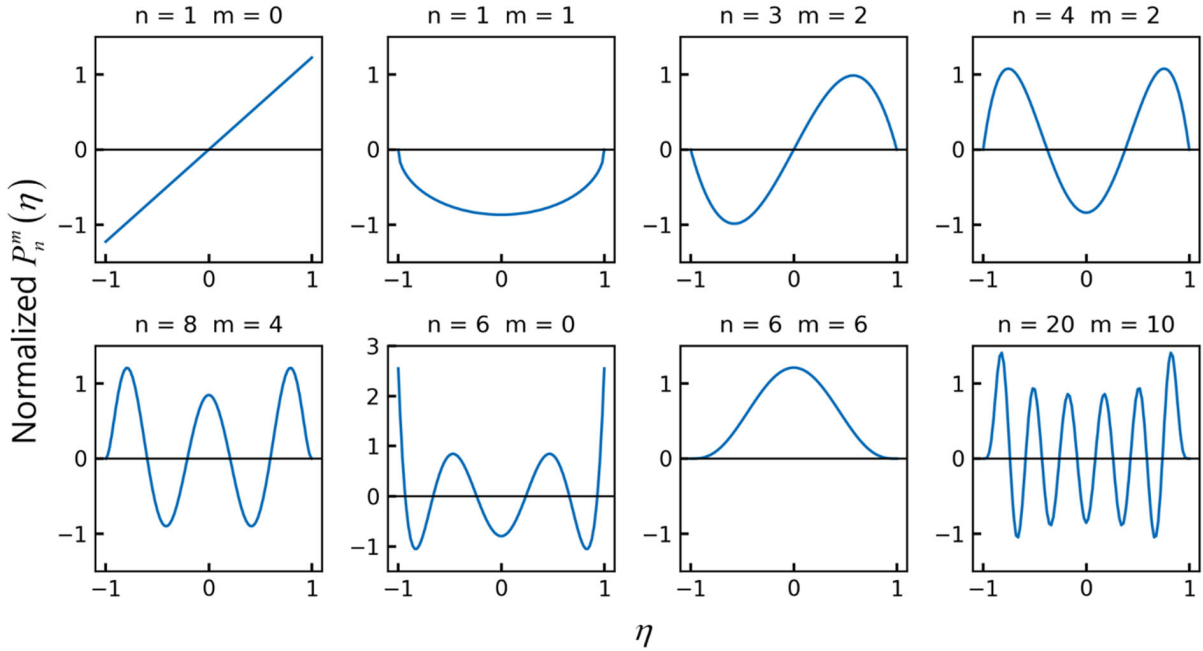

Fig. S2. Examples of normalized versions of associated Legendre functions for various values of $n$ and $m$.

We show some examples of these functions in Fig. S2. We have normalized these, following Eq. (53), by multiplying them by $\sqrt{(2n+1)(n-m)!/[2(n+m)!]}$. Note that these functions cross the zero axis $n-|m|$ times (the functions may or may not reach zero at the end points at $\eta = -1$ and $\eta = 1$, but they do not cross the zero axis there).

*Solving the $\phi$ equation*

The fact that $m$ has to be an integer means that we can write the solutions of the "$\phi$" equation (41) in a complex form

$$\Phi_m(\phi) = \exp(im\phi), \quad m = -n, \cdots, 0, \ldots n \tag{54}$$

or as two sets of real functions – odd "o" and even "e"

$$\Phi_{mo}(\phi) = \sin(m\phi), \quad m = 1, \ldots, n \tag{55}$$

$$\Phi_{me}(\phi) = \cos(m\phi), \quad m = 0, \ldots, n \tag{56}$$

(The omission of $m = 0$ for Eq. (55) is simply because in that case this function would be zero everywhere.) The choice of $m$ as an integer also means that these functions in Eqs. (54), (55), and (56) are single-valued in $\phi$. Whether one takes the "complex" or "real" function approach is a matter of taste and/or convenience for classical waves (though for quantum mechanical problems like the hydrogen atom, the complex form is required; see, e.g., [26]). Here, we use the "complex" approach, Eq. (54), for algebraic simplicity.

*Grouping the $\theta$ and $\phi$ solutions – spherical harmonics*

We can conveniently group the solutions to the "$\theta$" and "$\phi$" equations (48) and (41), respectively, into one function $Y$ of $\theta$ and $\phi$ with parameters $n$ and $m$ – the "spherical harmonic" functions (as in the forms in [25], p.108)

$$Y_{nm}(\theta,\phi) = \sqrt{\frac{(2n+1)}{4\pi}\frac{(n-m)!}{(n+m)!}} P_n^m(\cos\theta) \exp(im\phi) \tag{57}$$

which are normalized and orthogonal, i.e.,

$$\int_{\phi=0}^{2\pi}\int_{\theta=0}^{\pi} Y_{qp}^*(\theta,\phi) Y_{nm}(\theta,\phi) \sin\theta \, d\theta \, d\phi = \delta_{qn}\delta_{pm} \tag{58}$$

and, incidentally,

$$Y_{n,-m}(\theta,\phi) = (-1)^m Y_{nm}^*(\theta,\phi) \tag{59}$$

Spherical harmonics famously give the angular shapes of atomic orbitals in quantum mechanics and occur routinely in wave scattering from spherical objects and in many other classical problems, such as vibration modes of spherical shells. In hydrogen atom orbitals and in some other texts, the notation $l$ is used instead of $n$.

Though spherical harmonics may seem somewhat complicated functions, in fact they are relatively easy to visualize and describe. Spherical harmonics are simply functions of angle, so they can be conveniently plotted on the surface of a sphere, for example using color maps, as in Fig. S3. (The radius of the sphere has no particular meaning in such plots – the sphere is simply a convenient "canvas" on which to plot the spherical harmonic.)

Of course, to make such plots, we can plot only real numbers. We can usefully plot the real part of the spherical harmonics of Eq. (57), which correspond to the "even" functions of $\phi$ as given by Eq. (56). The "odd" functions (as in Eq. (55), which correspond to the imaginary part of the spherical harmonics of Eq. (57), have essentially the same shape, just rotated by an angle $\pi/2m$ around the polar axis. Note that, if we plot the spherical harmonics on a sphere of unit radius, the variable $\cos\theta$ used in the associated Legendre function $P_n^m(\cos\theta)$ in the definition of the spherical harmonics as in Eq. (57) can be viewed as the distance along the polar ($z$) axis from center of the sphere. (The example associated Legendre functions in Fig. S2 correspond to the spherical harmonics in Fig. S3.)

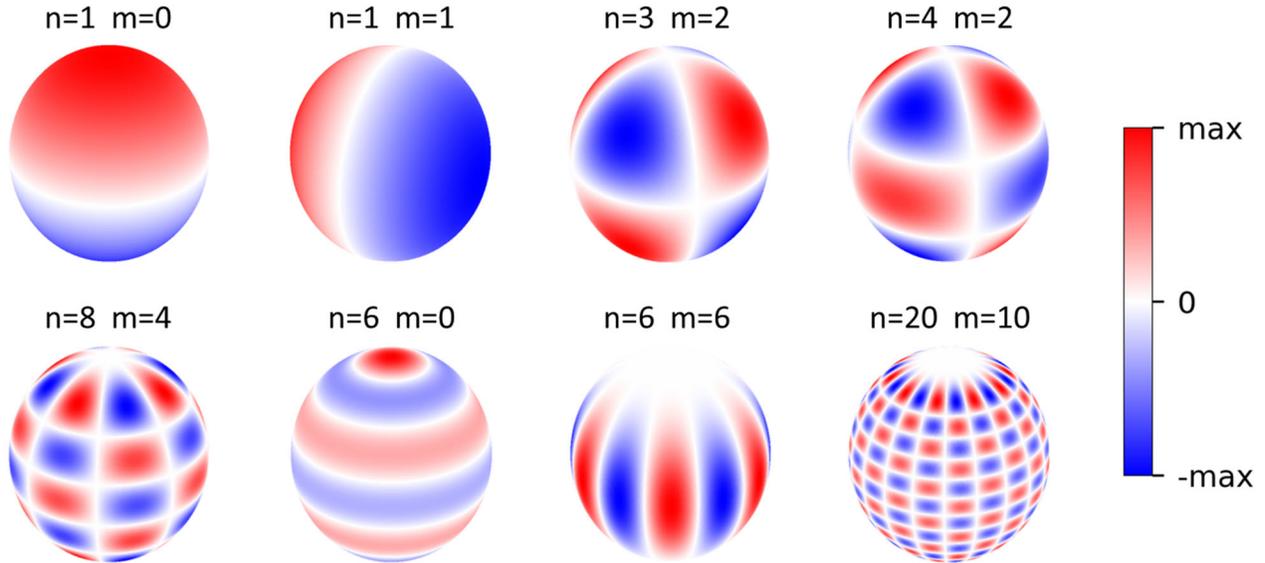

Fig. S3. Examples of spherical harmonics (technically the real part of the complex version of the functions), plotted as colormaps on a spherical surface. Spherical harmonics have $n$ nodal circles (i.e., circles on this spherical surface on which the function is identically zero), with $|m|$ such circles passing through both poles of the sphere. The remaining $n - |m|$ nodal circles are parallel to the equator on the sphere, positioned symmetrically with respect to the equator in the "northern" and "southern" hemispheres. For each plotted spherical harmonic, the color bar scale, shown on the right of the figure, runs between the largest magnitude, *max*, of the function at any angle (or position on this sphere), and -*max*. With this color bar scale, the nodal circles appear as white lines.

A simple way to describe the spherical harmonics (in their "real" forms as either sine or cosines in the angle $\phi$) is in terms of "nodal circles" (see, e.g., [26]). These are circles on the sphere on which the function is identically zero. They are either circles through both ("North" and "South") poles of the sphere (so, on lines of longitude on the sphere), or circles on or parallel to the equator (so, on lines of latitude on the sphere). The function always changes sign on passing through a nodal circle. The total number of such nodal circles is $n$. The number of (longitudinal) circles that pass through the poles is $m$ (as a positive number, so $|m|$ if one prefers), and there are $n - |m|$ ("latitudinal") nodal circles parallel to the equator.

The circles through the poles are always evenly spaced in azimuthal angle $\phi$. The latitudinal nodal circles parallel to the equator are always symmetrically spaced with respect to the equator. We could make a reasonable sketch of the spherical harmonics by using the equal spacing of the $|m|$ longitudinal nodal circles in $\phi$, and by placing the $n - |m|$ latitudinal nodal circles symmetrically with respect to the equator and reasonably evenly spaced in latitude. The precise positions in $\theta$ of the latitudinal nodal circles are given by the zeros of the corresponding associated Legendre function $P_n^m(\cos\theta)$ (which, of course, also has $n - |m|$ zero crossings as a function of $\theta$). Such pictures also correspond to the forms of the amplitude of the vibrations in the vibration modes of an actual hollow spherical shell ([31], pp. 1469-72) (at least for those vibrations that are "radial", corresponding to vibrations just "in and out" along a radius).

*Solving the radial equation – spherical Bessel functions*

The solutions to the radial equation (46) are the spherical Bessel functions of the first and second kinds, $j_n(\rho)$ and $y_n(\rho)$, respectively.

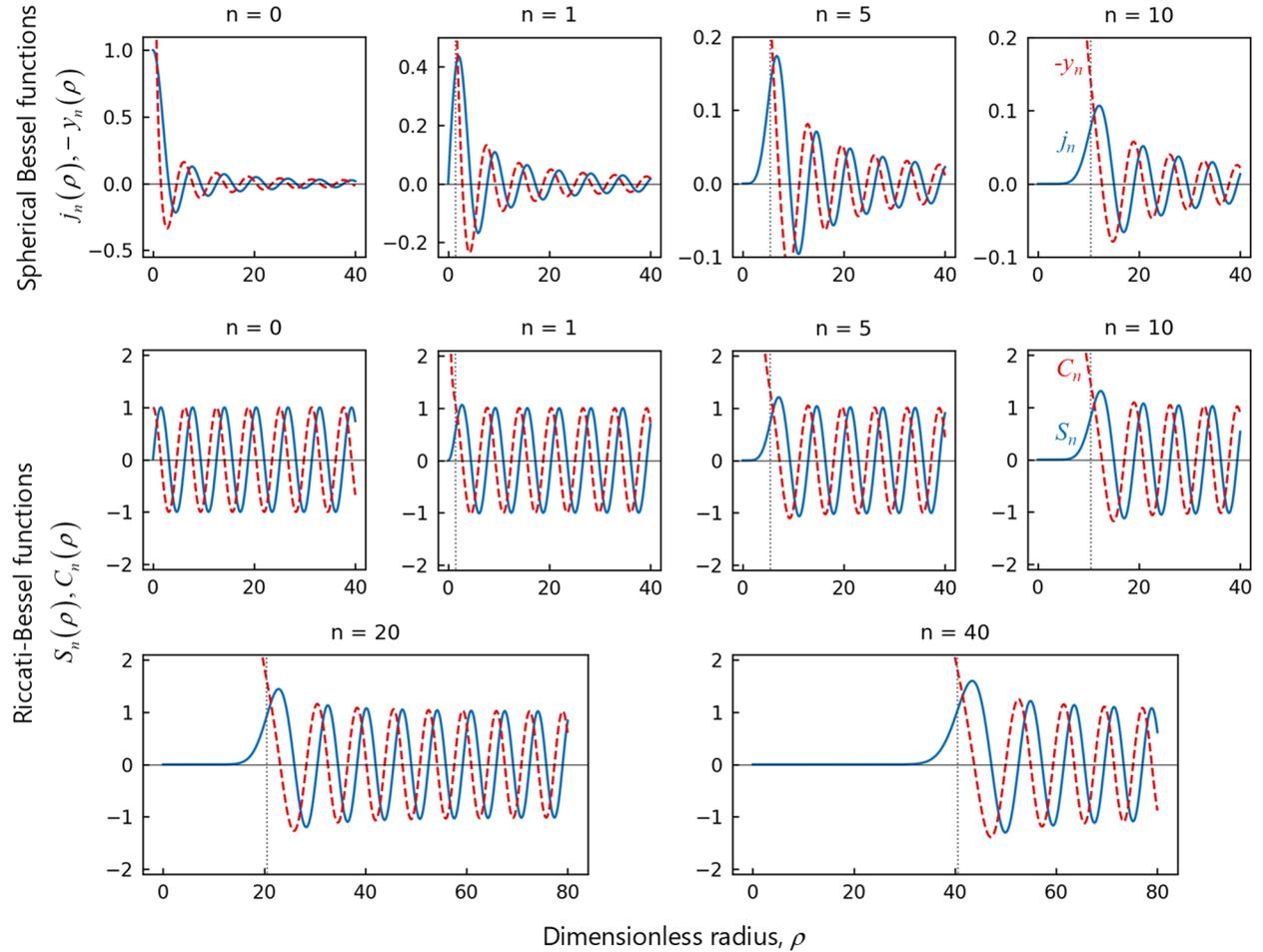

Fig. S4. Example spherical Bessel functions (top row) and Riccati-Bessel functions (middle and bottom rows). For the spherical Bessel functions, we plot $j_n(\rho)$ (solid blue lines) and $-y_n(\rho)$ (dashed red lines) (we use the minus in $-y_n(\rho)$ for graphic consistency with the Riccati-Bessel functions). For the Riccati-Bessel functions, we plot $S_n(\rho)$ (solid blue lines) and $C_n(\rho)$ (dashed red lines). All functions are plotted against the dimensionless "radius" variable $\rho$. The vertical dotted grey line represents the characteristic distance we call the escape radius $\rho_{escn} = \sqrt{n(n+1)}$

These are related to (ordinary) Bessel functions of half-integer order. Explicitly

$$j_n(\rho) = \sqrt{\frac{\pi}{2\rho}} J_{n+\frac{1}{2}}(\rho) \tag{60}$$

$$y_n(\rho) = \sqrt{\frac{\pi}{2\rho}} Y_{n+\frac{1}{2}}(\rho) \tag{61}$$

where $J$ and $Y$ are Bessel functions of the first and second kinds, respectively, and $n$ is conventionally taken to be zero or a positive integer.

Example spherical Bessel functions are graphed in the top row of Fig. S4. We have marked with a dotted grey line on each graph the characteristic radial distance that we call the "escape radius"

$$\rho_{escn} = \sqrt{n(n+1)} \tag{62}$$

in dimensionless form, or in dimensioned form

$$r_{escn} = \sqrt{n(n+1)}/k \tag{63}$$

From these formulas, we see immediately that

$$j_0(\rho) = \frac{\sin \rho}{\rho} \tag{64}$$

$$y_0(\rho) = -\frac{\cos \rho}{\rho} \tag{65}$$

Note that all the $y_n$ functions diverge at $\rho = 0$, and all the $j_n$ functions are zero at $\rho = 0$ except when $n = 0$, for which $j_0(0) = 1$.

It is also useful to look at the asymptotic behavior of $j_n(\rho)$ and $y_n(\rho)$ as $\rho \to \infty$, i.e., for large radius. Specifically (see, e.g., [25] p. 427)

$$j_n(\rho) \to \frac{1}{\rho} \sin\left(\rho - \frac{n\pi}{2}\right) \tag{66}$$

$$y_n(\rho) \to -\frac{1}{\rho} \cos\left(\rho - \frac{n\pi}{2}\right) \tag{67}$$

The solutions $j_n(\rho)$ and $y_n(\rho)$ correspond to standing waves in space. It is also useful to have propagating wave solutions. Since the equation of which these are the solutions, i.e., Eq. (46) (or equivalently Eq. (44)) is linear, linear superpositions of these solutions are also solutions. In particular, we can construct the solutions

$$h_n^{(1)}(\rho) = j_n(\rho) + i y_n(\rho) \tag{68}$$

$$h_n^{(2)}(\rho) = j_n(\rho) - i y_n(\rho) \tag{69}$$

$h_n^{(1)}(\rho)$ and $h_n^{(2)}(\rho)$ are known as the spherical Hankel functions of the first and second kinds, respectively. One reason for introducing them is that they can correspond to propagating waves. In particular, we see by substituting the asymptotic forms of $j_n(\rho)$, Eq. (66), and $y_n(\rho)$, Eq. (67), at large radius $\rho$ that

$$h_n^{(1)}(\rho) \to i\frac{\exp(i\rho)}{\rho} \equiv \frac{i}{k}\frac{\exp(ikr)}{r} \tag{70}$$

With our choice of $\exp(-i\omega t)$ as the oscillating temporal part of our waves as in Eq. (23), $h_n^{(1)}$ therefore corresponds at large radii to an outward propagating spherical wave. Similarly $h_n^{(2)}$ corresponds to an inward propagating spherical wave. It is a common notation to use $z_n(\rho)$ (as in Eqs. (44) and (46)) to refer to any of these solutions $j_n(\rho)$, $y_n(\rho)$, $h_n^{(1)}(\rho)$ or $h_n^{(2)}(\rho)$.

*Riccati-Bessel functions and differential equation*

At this point in other treatments of spherical waves (e.g., Refs. [25], [30]–[33], we would proceed to put the separated solutions together using Eq. (38), and we will do this shortly. However, there is one additional step we take here, which is central to our larger argument: we introduce the Riccati-Bessel functions and their differential equation.

We note first that, especially at large radial distances $\rho$, the spherical Bessel functions have an underlying "$1/\rho$" behavior. If we construct new functions where we take out that behavior by multiplying by $\rho$, then we may get a clearer picture of the behavior of the functions. So, we can construct the functions

$$S_n(\rho) = \rho j_n(\rho) \tag{71}$$

$$C_n(\rho) = -\rho y_n(\rho) \tag{72}$$

$$\xi_n(\rho) = \rho h_n^{(1)}(\rho) \equiv S_n(\rho) - iC_n(\rho) \tag{73}$$

$$\eta_n(\rho) = \rho h_n^{(2)}(\rho) \equiv S_n(\rho) + iC_n(\rho) \tag{74}$$

which are known as Riccati-Bessel functions, in various forms. In particular, the $\xi_n(\rho)$ function corresponds to an outward propagating spherical wave at large radii.

Note, incidentally, that $S_0(\rho) = \sin\rho$ and $C_0(\rho) = \cos\rho$, which explains the standard $S$ and $C$ notation choice. We have plotted example $S_n$ and $C_n$ functions in the second and third rows of Fig. S4. Note that, once we pass the escape radius, the functions converge quickly to unit amplitude of oscillation, which is equivalent to saying that the spherical Bessel functions converge quickly to underlying "$1/\rho$" behavior.

We can also regard the Riccati-Bessel functions as representing the wave as seen in angle rather than (transverse) position. For a spherically expanding wave of constant power, the modulus squared of the amplitude per unit solid angle does not change, consistent with the convergence of the Riccati-Bessel function amplitudes to unit magnitude oscillations with increasing radius.

Other than clarifying the behavior of the radial functions, these functions have another, and central, role in our work here. These Riccati-Bessel functions satisfy the differential equation

$$\rho^2 \frac{d^2\zeta_n}{d\rho^2} + \left(\rho^2 - n(n+1)\right)\zeta = 0 \tag{75}$$

where $\zeta_n$ refers to any of the functions $S_n$, $C_n$, $\xi_n$, or $\eta_n$. That this equation is correct is easily verified by substituting for $\zeta_n$ using any of the equations (71) – (74) and noting that

$$\frac{d^2}{d\rho^2}(\rho z_n) = \frac{d}{d\rho}\left[z_n + \rho\frac{dz_n}{d\rho}\right] = \rho\frac{d^2 z_n}{d\rho^2} + 2\frac{dz_n}{d\rho}$$

Using this substitution in Eq. (75) and dividing by $\rho$ recovers the spherical Bessel differential equation (46). The key point, though, is that we can rearrange the Riccati-Bessel equation (75) into the form

$$-\frac{d^2\zeta_n}{d\rho^2} + \frac{n(n+1)}{\rho^2}\zeta_n = \zeta_n \tag{76}$$

This is in the same form as a Schrödinger-like equation

$$-\frac{d^2\zeta_n}{d\rho^2} + V(\rho)\zeta_n = E_n\zeta_n \tag{77}$$

where the "eigenenergy" is $E_n = 1$ for all $n$, and in which the effective "potential" energy is $V(\rho) = n(n+1)/\rho^2$. So, for all radii $\rho$ for which the "potential" energy exceeds the "eigenenergy", we will have "tunneling" behavior – that is, for all $\rho$ for which

$$\frac{n(n+1)}{\rho^2} > 1 \tag{78}$$

or, equivalently,

$$\rho < \rho_{escn} = \sqrt{n(n+1)} \tag{79}$$

So, for $\rho < \rho_{escn}$, the wave is tunneling rather than propagating. Once $\rho$ exceeds the escape radius, the wave can start to propagate rather than tunnel. Fig. S5 illustrates the propagation of the resulting wave as function of time, for parameters as used for Fig. 1 of the main text, showing the outward propagation at larger radial distances.

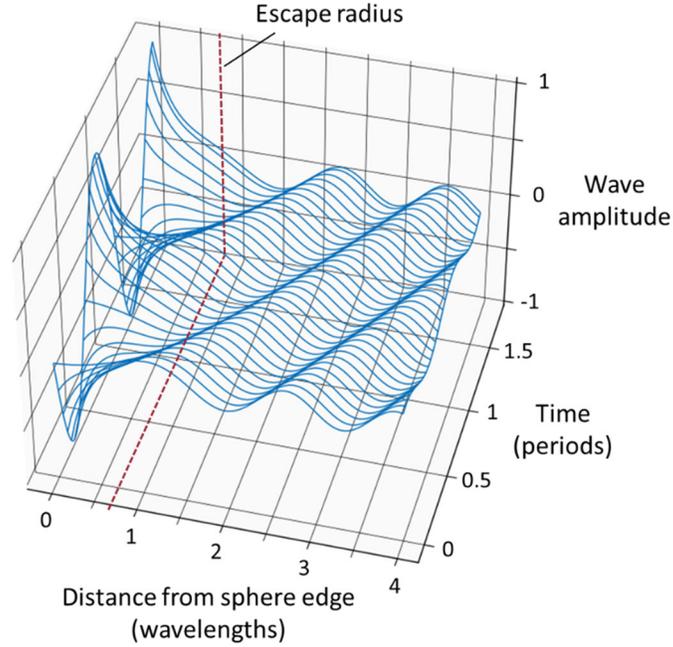

Fig. S5. Illustration of the propagation of the Riccati-Bessel wave, plotted here as the real part of $\xi_n(2\pi r)\exp(-2\pi i f_o t)$, where $r$ is the radial distance from the center of the spherical wave, $f_o$ is the frequency (so a period $t_o = 1/f_o$); 1.75 periods of oscillation are plotted. The wave has $n = 22$ and starts from a spherical surface of radius 2.9 wavelengths. The wave amplitude has been scaled to be 1 at the edge of the sphere. The escape radius is ~3.58 wavelengths (so ~0.68 wavelengths outside the spherical surface). These parameters are the same as used for Fig. 1 of the main text. The wave beyond the escape radius can be seen to propagate outwards as a function of time.

Expanding on the calculations in Figs. 1 and 2 of the main text, Fig. S6 shows the modulus squared of the relative far field magnitude of the spherically expanding wave $\gamma_n(kr_o) = |\xi_n(\infty)|^2 / |\xi_n(kr_o)|^2$ (Eq. (19) of the main text) for various radii $r_o$ of the spherical surface, as a function of the order $n$ of the spherical wave and, parametrically, of the total cumulative number of spherical waves up to and including order $n$ (here $(n+1)^2$ for the scalar wave case). This shows how the tunneling "tail" of this coupling decreases as the radius of the spherical surface increases.

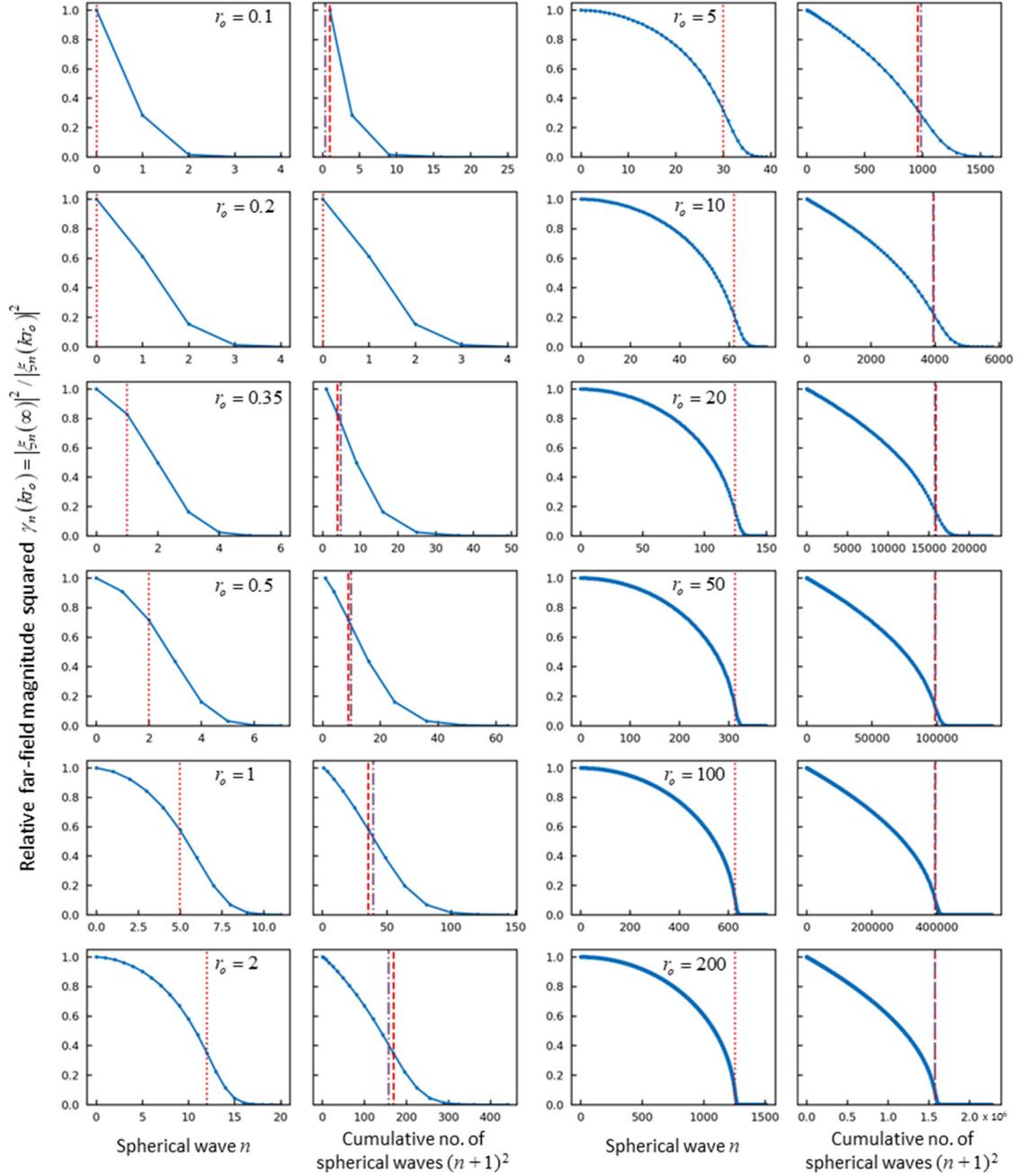

Fig. S6. Extended plots as in Fig. 2, here for the scalar wave case including $n=0$. Relative far-field magnitude squared $\gamma_n(kr_o)$ of the spherical wave with increasing $n$ for various values of the starting radius $r_o$ (in wavelengths) of the spherical surface. Lines are just to guide the eye. $n_p$ (vertical dotted lines), $N_{ps} = (n_p+1)^2$ (vertical dashed lines), and the spherical heuristic number $N_{SH} = (kr_o)^2$ (vertical dash-dot lines) are shown for each case, similar to Fig. 2.

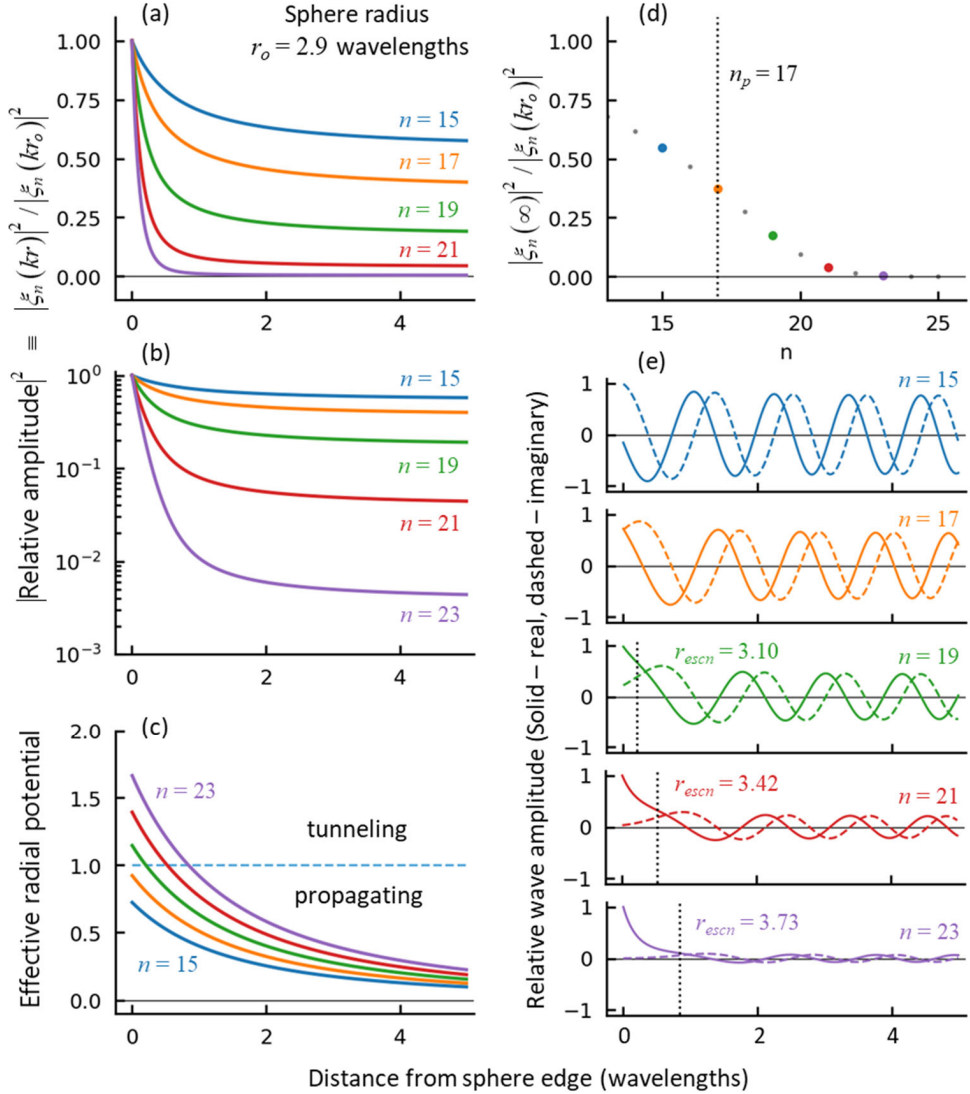

Fig. S7. Effective radial potentials and waves starting from a spherical surface of radius $r_o = 2.9$ wavelengths. Curves are plotted for $n = 15, 17, 19, 21,$ and $23$, as a function of radial distance from the edge of the sphere. ($r$ is the distance from the center of the spherical surface.) (a) and (b) Relative magnitude squared of the wave as a function of radial distance from the spherical surface, on (a) linear and (b) log scales. (c) Effective radial potential $n(n+1)/(kr)^2$. The dashed line corresponds to the division between tunneling and propagating behavior. (d) Relative magnitude squared of the wave at infinite distance as a function of $n$. (This is an expanded version of Fig. 2(b), with specific $n$ values indicated.) (e) Real and imaginary parts of the wave $i\xi_n(kr) = C_n(kr) + iS_n(kr)$ for the different $n$ values, normalized to unit magnitude at the sphere edge in each case. The escape radius $r_{escn}$ (in wavelengths) is also shown (vertical dotted lines). For $n = 15$ and $17$ the escape radii of 2.47 and 2.78 wavelengths are less than $r_o$, so these radii are not shown on these plots (these waves are always propagating).

Fig. S7 shows detailed behavior of the Riccati-Bessel waves for different *n* for the example case of $r_o = 2.9$ wavelengths, showing the progression with increasing *n* from waves that are initially propagating to ones requiring tunneling escape. The tunneling-like behavior up to the escape radius $r_{escn}$ is quite clear by eye in, for example, Fig. S7(e) for the $n = 21$ and $n = 23$ curves.

Curves similar to Fig. S7 but for other radii of the spherical surface are given in Figs. S8, S9, S10, S11 and S12, showing these behaviors and their consequences. When the radius $r_o$ and the corresponding *n* for the first few tunneling waves are small numbers, the tunneling barrier, as given by $V(kr) = n(n+1)/(kr)^2$, Eq. (8), for the first several "escaping" waves tends to be high but thin. For large $r_o$ and corresponding *n*, the barrier tends to be relatively low but thick for those first "escaping" waves.

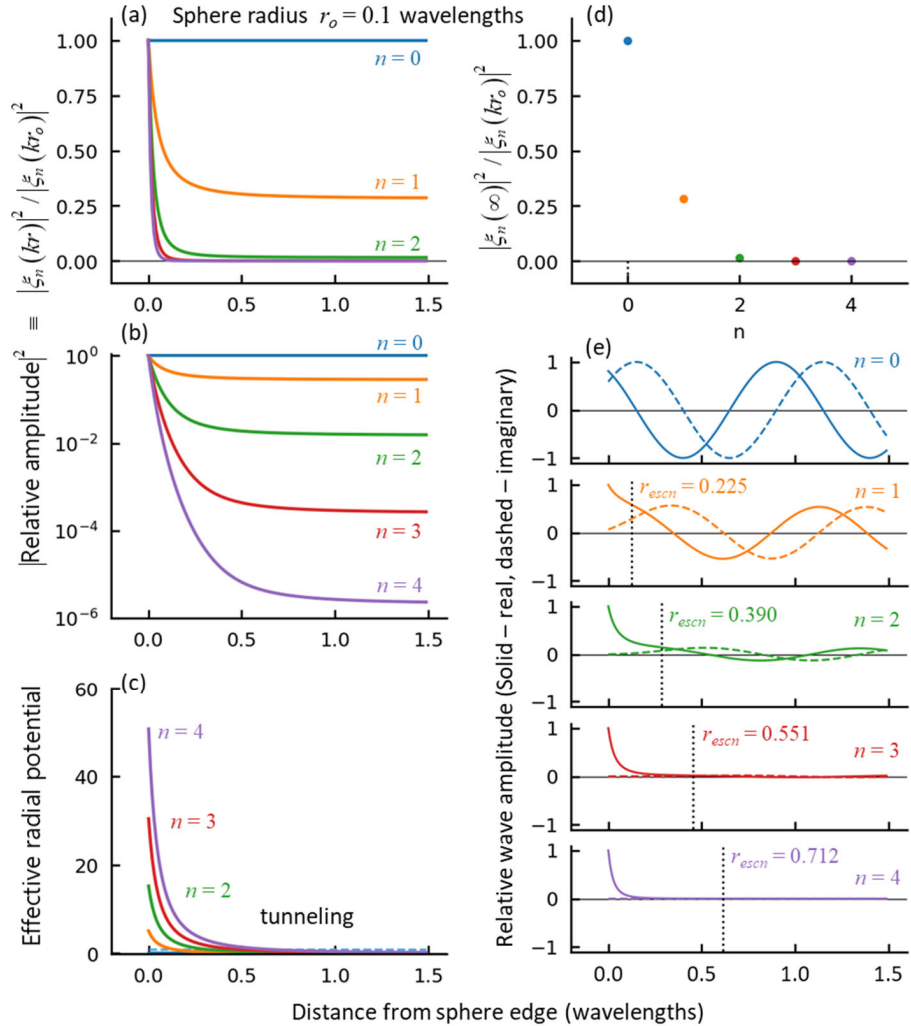

Fig. S8. Effective radial potentials and waves as in Fig. S7 but starting from a spherical surface of radius $r_o = 0.1$ wavelengths. Curves are plotted for $n = 0, 1, 2, 3,$ and 4, as a function of radial distance from the edge of the sphere.

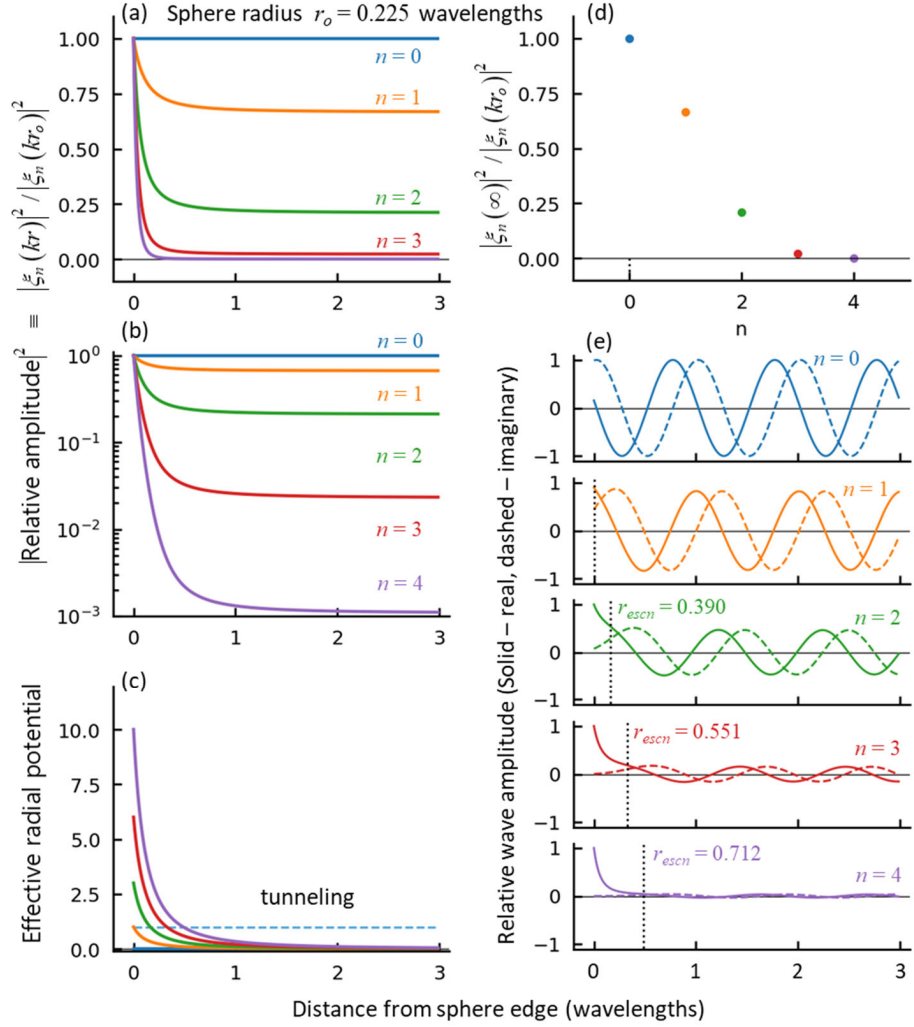

Fig. S9. Effective radial potentials and waves as in Fig. S7 but starting from a spherical surface of radius $r_o = r_{esc1} = \lambda_o / (\sqrt{2}\,\pi) \simeq 0.225\lambda_o$ wavelengths (see Eq. (16)). Curves are plotted for $n$ = 0, 1, 2, 3, and 4.

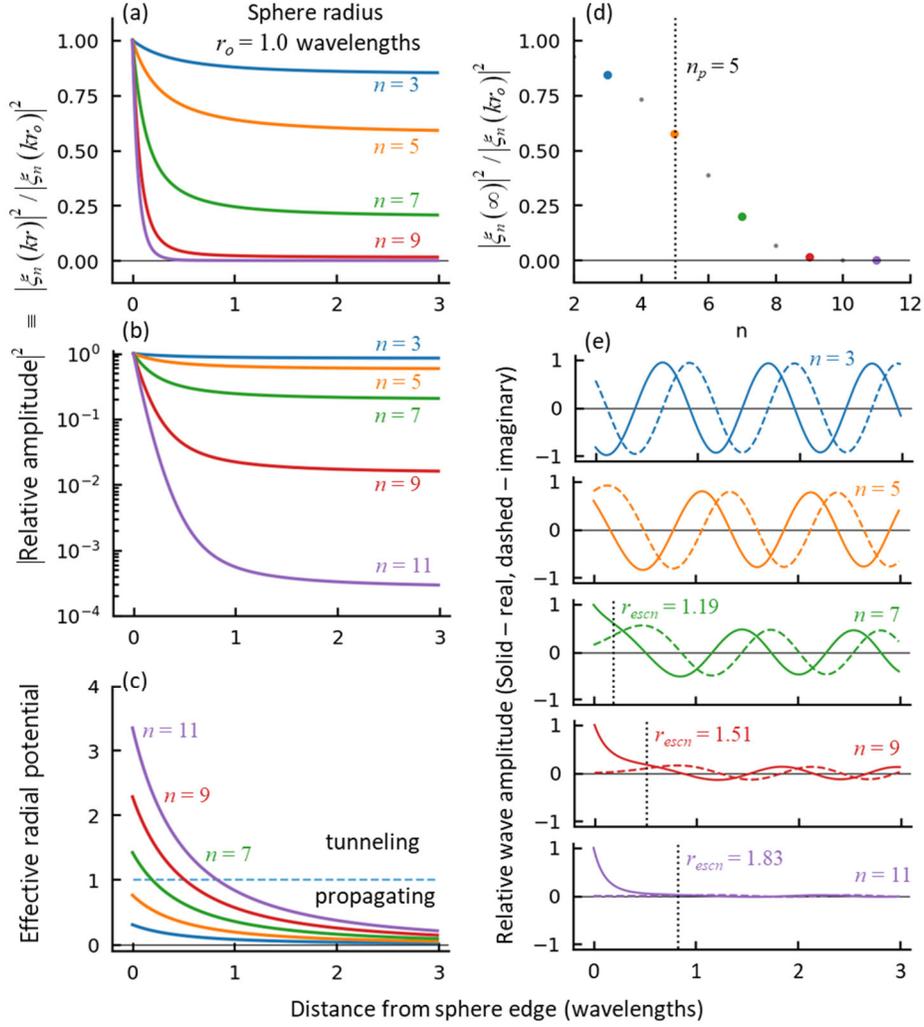

Fig. S10. Effective radial potentials and waves as in Fig. S7 but starting from a spherical surface of radius $r_o = 1.0$ wavelengths. Curves are plotted for $n$ = 3, 5, 7, 9, and 11. The escape radius $r_{escn}$ (in wavelengths) is also shown (vertical dotted lines). For $n = 3$ and $n = 5$, the escape radii of 0.551 and 0.872 wavelengths, respectively, are less than $r_o$, so they do not appear on this plot (these waves are always propagating).

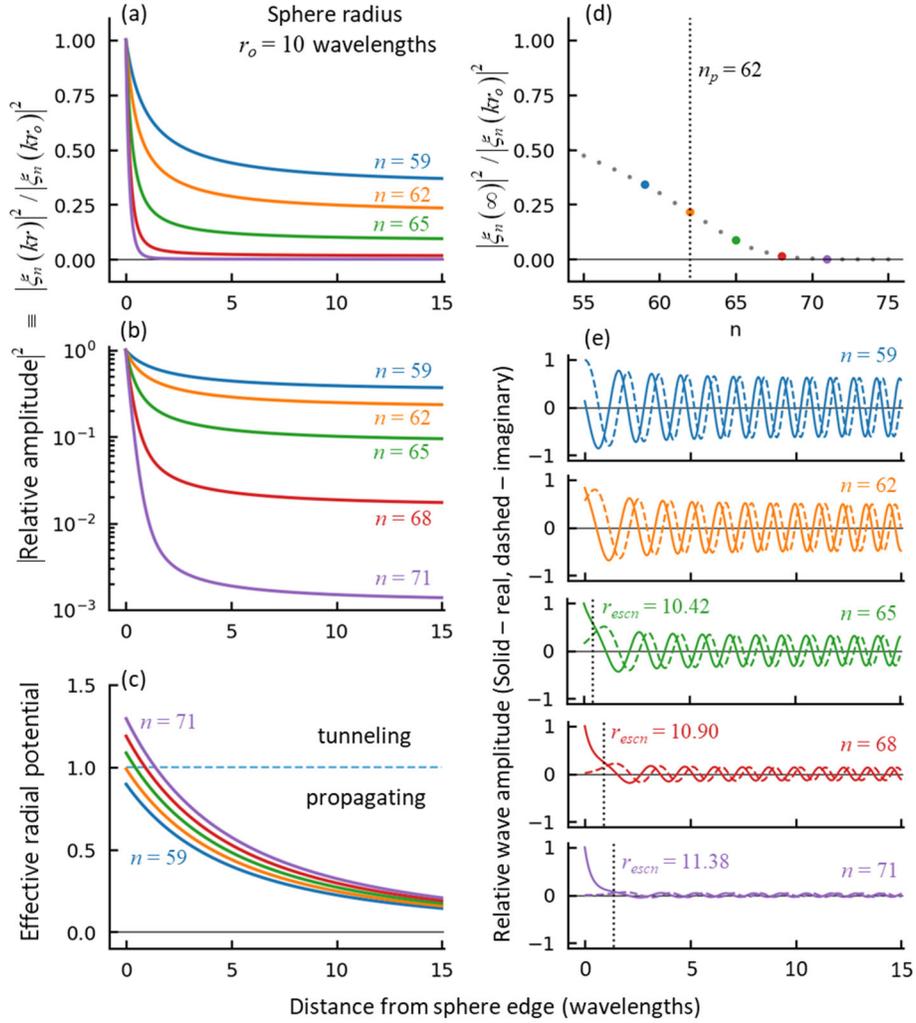

Fig. S11. Effective radial potentials and waves as in Fig. S7 but starting from a spherical surface of radius $r_o = 10$ wavelengths. Curves are plotted for $n$ = 59, 62, 65, 68, and 71. The escape radius $r_{escn}$ (in wavelengths) is also shown (vertical dotted lines). For $n = 59$ and $n = 62$, the escape radii of 9.47 and 9.95 wavelengths, respectively, are less than $r_o$, so they do not appear on this plot (these waves are always propagating).

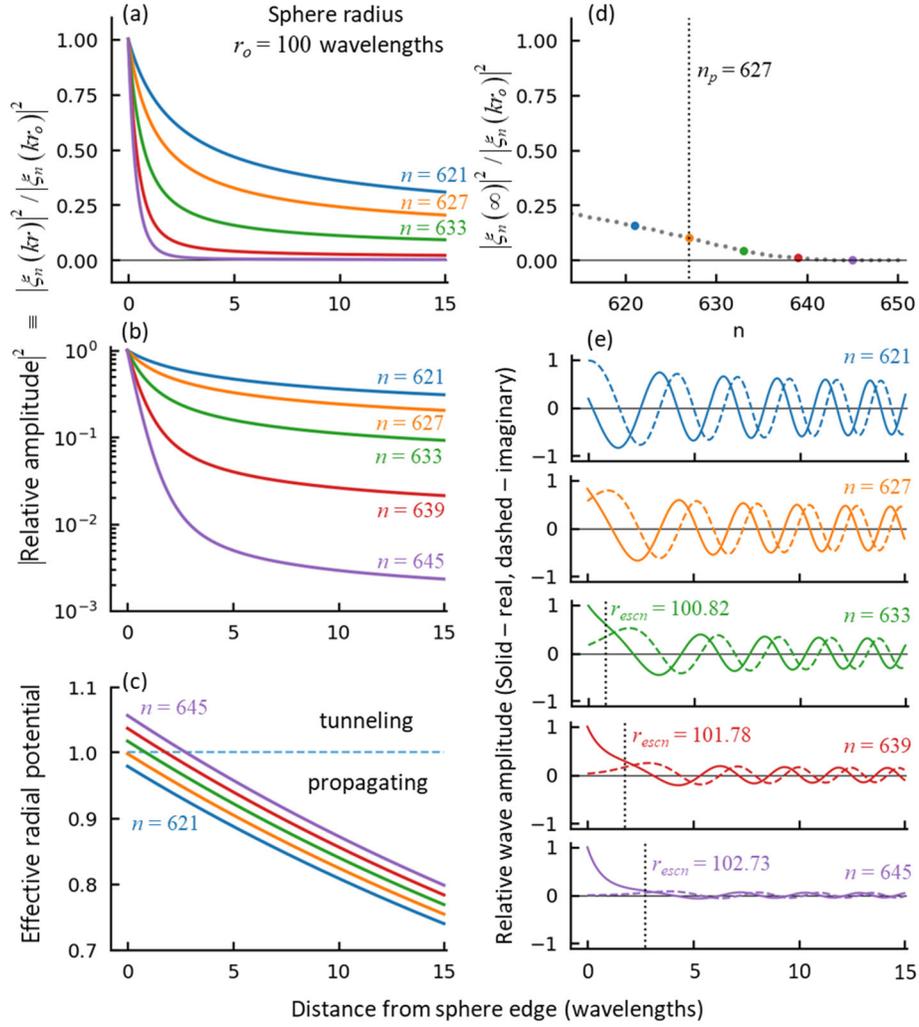

Fig. S12. Effective radial potentials and waves as in Fig. S7 but starting from a spherical surface of radius $r_o = 100$ wavelengths. Curves are plotted for $n$ = 621, 627, 633, 639, and 645. For $n = 621$ and $n = 627$, the escape radii of 98.91 and 99.87 wavelengths, respectively, are less than $r_o$, so they do not appear on this plot (these waves are always propagating).

*Change of character of the solutions at the escape radius*

Note from Eq.(77) that, at this transition from tunneling to propagating (so $V = E$), the second derivative of the Riccati-Bessel functions is exactly zero. (Indeed, if one prefers, one can use this mathematical definition instead of the Schrödinger equation analogy to describe the escape radius and the wave behavior.) To be more specific, from Eq. (77), at this transition (i.e., at the escape radius $r_{escn}$, or, equivalently, $\rho_{escn}$)

$$\frac{1}{\zeta_n(\rho)} \frac{d^2 \zeta_n(\rho)}{d\rho^2} \bigg|_{\rho = \rho_{escn}} = 0 \tag{80}$$

This value $\rho_{escn}$ is the only radius for which this is the case for both the real and imaginary parts of $\zeta_n$, and this can be taken as an alternative and equivalent mathematical way of defining the escape radius.

For smaller radii (so, $V > E$), this derivative (in the form $(1/\zeta_n)(d^2\zeta_n/d\rho^2)$) is always positive, so these functions are never oscillatory for radii below the escape radius. Indeed, for some function $f(s)$ as a function of some position variable $s$, we could regard the criterion

$$\frac{1}{f} \frac{d^2 f}{ds^2} > 0 \tag{81}$$

as defining what we mean by "quasi-exponential" or "tunneling" and the complementary criterion

$$\frac{1}{f} \frac{d^2 f}{ds^2} < 0 \tag{82}$$

as what we mean by "quasi-oscillatory" or "propagating".

Once we pass the escape radius (so, $E > V$), $(1/\zeta_n)(d^2\zeta_n/d\rho^2)$ is always negative. Then, when the function $\zeta_n$ is positive, the second derivative is negative, so the function eventually must turn "down", which eventually pushes the function negative. When the function $\zeta_n$ is negative, the second derivative is then positive, which means the function eventually must turn "up", which pushes the function positive, and so on, giving some kind of oscillating behavior. This is consistent with the graphs in Fig. S4, and we may also be able to discern by eye the change in character in the Riccati-Bessel functions from "quasi-exponential" to "quasi-oscillatory" as the radius passes this escape radius.

Again, note that this transition from tunneling to propagating behavior, though clear from our discussion here of the differential equation (76), is not at all obvious from the explicit formulas for Riccati-Bessel functions in terms of sines, cosines and series of inverse powers of the radius (see, e.g., [25], p. 426). For example, the Riccati-Bessel functions for $n = 1$ have explicit formulas

$$S_1(\rho) = \frac{\sin \rho}{\rho} - \cos \rho, \ C_1(\rho) = \frac{-\cos \rho}{\rho} - \sin \rho \tag{83}$$

It is straightforward to verify algebraically that the second derivative of both of these functions is zero when $\rho = \sqrt{2}$ (which is the corresponding escape radius in dimensionless units for the outward Riccati-Bessel function $\xi_1(\rho)$), but it is far from obvious from these functional forms that we have tunneling behavior for $\rho < \sqrt{2}$ and propagating behavior for $\rho > \sqrt{2}$.

## S1.3 Complete solutions for scalar spherical waves

Now we can return to the original proposal for separable solutions in Eq. (38). With the spherical harmonics $Y_{nm}(\theta,\phi)$ as a convenient way of writing the functions $\Theta(\theta)\Phi(\phi)$, and with $z_n(\rho) = z_n(kr)$ representing the spherical Bessel functions in any of the forms $j_n(\rho)$, $y_n(\rho)$, $h_n^{(1)}(\rho)$ or $h_n^{(2)}(\rho)$, and similarly $\zeta_n(\rho) = \zeta_n(kr)$ representing the Riccati-Bessel functions in any of the forms $S_n(\rho)$, $C_n(\rho)$, $\xi_n(\rho)$, or $\eta_n(\rho)$, the separable solutions of the scalar Helmholtz equation (24) in spherical polar coordinates are given by

$$U_{nm}(\mathbf{r}) = z_n(kr) Y_{nm}(\theta,\phi) \equiv \frac{1}{kr}\zeta_n(kr) Y_{nm}(\theta,\phi) \tag{84}$$

A key point about such a separated solution is that, for any such wave, it retains the same angular shape $Y_{nm}(\theta,\phi)$ at all radii $r$ or $\rho$. Note, incidentally, that beams that retain the form of their shape as they propagate are unusual in solutions of wave equations, especially ones that make no "paraxial" approximations. (Hermite-Gaussian or Laguerre-Gaussian beams in free-space can retain the form of their shape, with a scale factor, as they propagate, but they are based on paraxial approximations.)

We are particularly interested in the waves that are outward-going waves at large radii – essentially, we are imposing a "radiation condition" on our solutions. In that case, the possible separated solutions are

$$U_{nmout}(\mathbf{r}) = h_n^{(1)}(kr) Y_{nm}(\theta,\phi) \equiv \frac{1}{kr}\xi_n(kr) Y_{nm}(\theta,\phi) \tag{85}$$

## S1.4 Representing arbitrary outgoing scalar waves in spherical waves

Now, spherical harmonics form a complete set for describing any function $f_{ro}(\theta,\phi)$ of the angles $\theta$ and $\phi$ (see, e.g., Ref. [25], p. 108). Formally, we could decompose $f_{ro}(\theta,\phi)$ onto the spherical harmonic basis, writing

$$f_{ro}(\theta,\phi) = \sum_{n=0}^{\infty} \sum_{m=-n}^{n} a_{nm} Y_{nm}(\theta,\phi) \tag{86}$$

where

$$a_{nm} = \int_{\phi=0}^{2\pi} \int_{\theta=0}^{\pi} Y_{nm}^*(\theta,\phi) f_{ro}(\theta,\phi) \sin\theta\, d\theta\, d\phi \tag{87}$$

With these coefficients, we can now represent the entire outgoing wave $\psi(r,\theta,\phi)$. Formally, that expansion takes the form

$$\psi(r,\theta,\phi) = \sum_{n=0}^{\infty} \sum_{m=-n}^{n} b_{nm} h_n^{(1)}(kr) Y_{nm}(\theta,\phi) \tag{88}$$

for some expansion coefficients $b_{nm}$. Since we presume we know the angular form $f_{ro}(\theta,\phi)$ at radius $r_o$, then we have

$$\psi(r_o,\theta,\phi) = f_{ro}(\theta,\phi) = \sum_{n=0}^{\infty} \sum_{m=-n}^{n} a_{nm} Y_{nm}(\theta,\phi) \equiv \sum_{n=0}^{\infty} \sum_{m=-n}^{n} b_{nm} h_n^{(1)}(kr_o) Y_{nm}(\theta,\phi) \tag{89}$$

from which we can conclude that

$$b_{nm} h_n^{(1)}(kr_o) = a_{nm} \tag{90}$$

or, equivalently,

$$b_{nm} = a_{nm} / h_n^{(1)}(kr_o) \tag{91}$$

Hence, given the angular form $f_{ro}(\theta,\phi)$ of an outgoing wave at any particular radius $r_o$, we can expand it in the set of functions $U_{nmout}(\mathbf{r})$ as in Eq. (85), and describe this wave at all other positions in this empty space or uniform medium outside this spherical surface of radius $r_o$.

Of course, it is *not* the case that such an arbitrary outgoing wave $\psi(r,\theta,\phi)$ generally remains the same angular shape as it propagates. The specific functions $U_{nmout}(\mathbf{r})$ or, since the orientation of axes is arbitrary, any rotated version of them, do, however, have this property of retaining their shape as they propagate out. It is also the case that any linear combination of such functions with the same *n* but different *m* values will have this property because they all share the same radial function $h_1^{(1)}(r)$.

## S2 Vector waves in spherical coordinates

Vector wave solutions in spherical coordinates have been studied extensively, starting at least with Mie[19] and Debye [20], especially to understand scattering properties of spherical particles and other objects, with significant current technological interest also[22], [34]. Textbook treatments are given, e.g., by Stratton [30], Morse and Feshbach[35], Bohren and Huffman[32], and Tsang et al.[33]. Our primary interest is in counting orthogonal waves rather than solving scattering problems. Since our results relate to different phenomena, especially tunneling escape of waves, we give a relatively complete discussion and derivation of the major results here.

Based on an approach by Bouwcamp and Casimir[36], there is a relatively short derivation that applies only to vector electromagnetic fields – explicitly the electric and magnetic fields **E** and **H**. This approach is based on the scalar fields $\mathbf{r} \cdot \mathbf{E}$ and $\mathbf{r} \cdot \mathbf{H}$, which must[25], [36] also satisfy the scalar Helmholtz equation, Eq. (24); this is the approach used by Jackson[25] to derive the solutions (and specifically the multipole field expansion) for vector electromagnetic fields. Since we want also formally to prove that the "tunneling" analysis can apply to other vector fields as well, including acoustic and elastic waves, we give the more general derivation here.

We start by defining some general mathematics for vector fields in spherical coordinates. Initially, we postpone discussion of the special case of electromagnetic fields.

## S2.1 Vector spherical harmonics

As a preliminary, we define the vector spherical harmonic functions, [35], p. 1898-99 and [33], p. 27). These are three sets of vector functions of the spherical polar coordinates $(r,\theta,\phi)$ that can describe any vector function of the angles $\theta$ and $\phi$ on a spherical surface of radius $r$. They are based on the spherical harmonic functions $Y_{nm}(\theta,\phi)$ and some of their vector derivatives. They can be defined[33] as

$$\mathbf{P}_{mn}(\theta,\phi) = \hat{\mathbf{r}} Y_{nm}(\theta,\phi) \quad n = 0,1,2..., \quad -n \leq m \leq n \tag{92}$$

$$\mathbf{B}_{mn}(\theta,\phi) = r\nabla Y_{nm}(\theta,\phi) \quad n = 1,2..., \quad -n \leq m \leq n \tag{93}$$

$$\mathbf{C}_{mn}(\theta,\phi) = \nabla \times \left[\mathbf{r} Y_{nm}(\theta,\phi)\right] \quad n = 1,2..., \quad -n \leq m \leq n \tag{94}$$

where $\hat{\mathbf{r}}$ is a unit vector in the radial direction for the point of interest on the spherical surface.

These different functions are all orthogonal to one another in the angles $\theta$ and $\phi$, both for different $n$ and/or $m$ and, for different ones of the $\mathbf{P}_{mn}$, $\mathbf{B}_{mn}$, and $\mathbf{C}_{mn}$, even for the same $n$ and $m$. Taken together, they form complete sets for describing any vector field as a function of angles $\theta$ and $\phi$ on a spherical surface of radius $r$. Note that $\mathbf{P}_{mn}$ describes functions that are radially polarized (or "longitudinal"), i.e., vectors that point exactly inwards or outwards, perpendicular to the spherical surface.

In component form, we can perform the vector derivatives for $\mathbf{B}_{mn}$ and $\mathbf{C}_{mn}$. First,

$$\mathbf{B}_{mn}(\theta,\phi) = \left[\hat{\theta}\frac{d}{d\theta}P_n^m(\cos\theta) + \hat{\phi}\frac{im}{\sin\theta}P_n^m(\cos\theta)\right]\exp(im\phi) \tag{95}$$

For $\mathbf{C}_{mn}$, we note that, because of the vector calculus identity

$$\nabla \times (f\mathbf{a}) = \nabla f \times \mathbf{a} + f\nabla \times \mathbf{a} \tag{96}$$

and the fact ([37], pp. 149-150) that

$$\nabla \times \mathbf{r} = 0 \tag{97}$$

we can rewrite Eq. (94) as

$$\mathbf{C}_{mn}(\theta,\phi) = \nabla Y_{nm}(\theta,\phi) \times \mathbf{r} \quad (\equiv -\hat{\mathbf{r}} \times \mathbf{B}_{mn}(\theta,\phi)) \tag{98}$$

and so we obtain

$$\mathbf{C}_{mn}(\theta,\phi) = \left[\hat{\theta}\frac{im}{\sin\theta}P_n^m(\cos\theta) - \hat{\phi}\frac{d}{d\theta}P_n^m(\cos\theta)\right]\exp(im\phi) \tag{99}$$

Incidentally, note also the relation

$$\mathbf{B}_{mn}(\theta,\phi) = \hat{\mathbf{r}} \times \mathbf{C}_{mn}(\theta,\phi) \tag{100}$$

which is readily proved from these component forms.

Eqs. (100) and (98) imply that both $\mathbf{B}_{mn}$ and $\mathbf{C}_{mn}$ represent fields that are tangential to the spherical surface, so they are both expressible entirely in terms of components in the $\hat{\theta}$ and $\hat{\phi}$ directions, i.e., they represent "transverse" fields on the spherical surface. (Note: $\mathbf{B}_{mn}$ should not be confused with the $\mathbf{B}$ magnetic field; here it represents just a specific mathematical vector function related to spherical harmonics.)

Note above that, while in Eq. (92) we have listed $\mathbf{P}_{mn}$ as being defined for the integer $n$ starting from 0, for $\mathbf{B}_{mn}$ and $\mathbf{C}_{mn}$, $n$ starts from 1. This turns out to be quite important for the behavior of vector waves, especially electromagnetic ones. We can see the reason for this if we remember that the spherical harmonic for $n=0$, $Y_{00}$, is the same for all angles $\theta$ and $\phi$. Since spherical harmonics do not depend on radius $r$, being only functions of angle, then all derivatives of $Y_{00}$ are zero, so $\nabla Y_{00}$ is the zero vector. From the definition Eq. (93) that means $\mathbf{B}_{00}$ would also be the zero vector. Similarly, from the definition Eq. (98), it follows that $\mathbf{C}_{00}$ would also be zero.

Note also that these various vector spherical harmonics represent orthogonal polarization directions. Obviously, $\mathbf{P}_{mn}$ represents only radial polarized fields (i.e., in the $\hat{\mathbf{r}}$ direction), and neither $\mathbf{B}_{mn}$ nor $\mathbf{C}_{mn}$ has any radial components, so they are fields that are always tangential to a spherical surface. Also, we can verify directly from the component forms Eqs. (95) and (99) that

$$\mathbf{B}_{mn} \cdot \mathbf{C}_{mn} = 0 \tag{101}$$

so $\mathbf{B}_{mn}$ and $\mathbf{C}_{mn}$ represent orthogonal polarizations on the spherical surface.

## S2.2 General vector fields in spherical coordinates

Consider any vector field $\mathbf{F}$ in three-dimensional space that it is sufficiently differentiable and falls off at least as fast as $1/r$ as $r \to \infty$, where $r$ is the radius of some large bounding sphere. Then any such field $\mathbf{F}$ can be decomposed into two parts

$$\mathbf{F} = \mathbf{L} + \mathbf{T} \tag{102}$$

where

$$\mathbf{L} = \nabla \phi_L \tag{103}$$

is the "longitudinal" field, where $\phi_L$ is a scalar function of space (which we call a scalar potential), and

$$\mathbf{T} = \nabla \times \mathbf{A}_T \tag{104}$$

is the "transverse" field and $\mathbf{A}_T$ is a vector function of space, which we call a vector potential. (Note we are not yet identifying $\phi_L$ or $\mathbf{A}_T$ with any scalar or vector potentials in electromagnetism; this is a general statement for such fields. Note too that $\mathbf{L}$ should not be confused with other uses of this notation for angular momentum.) This decomposition (102) is known as Helmholtz's theorem or the Helmholtz decomposition (see, e.g., [35], p. 1763 and [37], p. 178).

Though **L** and **T** are referred to as "longitudinal" and "transverse", **L** is not necessarily a purely radially polarized field in spherical coordinates, nor is **T** necessarily polarized transverse to the radius vector in spherical coordinates, though they can be purely longitudinal and transverse respectively at large distances (i.e., in the "far field") from the center of some source. Nonetheless, these "longitudinal" and "transverse" notations are common and convenient.

Now let us presume that our field **F** of interest is to satisfy the vector Helmholtz equation

$$\nabla^2 \mathbf{F} + k^2 \mathbf{F} \equiv \nabla(\nabla \cdot \mathbf{F}) - \nabla \times \nabla \times \mathbf{F} + k^2 \mathbf{F} = 0 \tag{105}$$

where we use the vector calculus identity

$$\nabla^2 \mathbf{f} \equiv \nabla(\nabla \cdot \mathbf{f}) - \nabla \times \nabla \times \mathbf{f} \tag{106}$$

for any appropriately differentiable vector function **f**. (We will formally prove below that electromagnetic waves satisfy the vector Helmholtz equation, though for the moment we just presume we have some vector field that satisfies this equation.)

We consider first the longitudinal field $\mathbf{L} = \nabla \phi_L$. (We do this mostly for completeness; this component will not appear for electromagnetic fields at finite frequencies, as we will show below, though it may exist for elastic waves, for example as longitudinal sound waves in solid media.). Noting that, for any scalar function $f$, $\nabla \times \nabla f = 0$ as a vector calculus identity, then Eq. (105) becomes

$$\nabla(\nabla \cdot \mathbf{L}) + k^2 \mathbf{L} = 0 \tag{107}$$

So, noting that $\nabla \cdot \nabla f \equiv \nabla^2 f$ by definition, then substituting from Eq. (103) in Eq. (107)

$$-\nabla(\nabla^2 \phi_L + k^2 \phi_L) = 0 \tag{108}$$

So **L** is a solution of the vector Helmholtz equation if $\phi_L$ is a solution of the scalar Helmholtz equation

$$\nabla^2 \phi_L + k^2 \phi_L = 0 \tag{109}$$

(so longitudinal sound waves, modeled using an underlying scalar Helmholtz equation, can correspond to solutions of the vector Helmholtz equation, for example).

Now we consider the "transverse" components of the fields. Quite generally, since $\nabla \cdot (\nabla \times \mathbf{f}) = 0$ for any appropriately differentiable vector function **f**, then from Eq. (104) we have $\nabla \cdot \mathbf{T} = 0$. Hence, for such "transverse" fields, the vector Helmholtz equation, Eq. (105), becomes

$$\nabla \times \nabla \times \mathbf{T} = k^2 \mathbf{T} \tag{110}$$

Following Stratton[30], pp. 414-5 (see also [35], pp. 1764-6), we can propose a solution for Eq. (110) in the form

$$\mathbf{M} = \nabla \times (\mathbf{r} \psi_M) \tag{111}$$

where **r** is the position vector of the point of interest, which we can regard as a radial vector in spherical polar coordinates. (So, we are choosing a "vector potential" function $\mathbf{A}_F = \mathbf{r}\psi_M$ as in Eq. (104); note we are *not* identifying this as the magnetic vector potential of electromagnetism – this is just a mathematical construct.)

Using the identities Eqs. (96) and (97), we can rewrite Eq. (111) as

$$\mathbf{M} = \nabla \psi_M \times \mathbf{r} \tag{112}$$

which means **M** is perpendicular to **r**, so there is no component of **M** in the $\hat{\mathbf{r}}$ direction. Hence, performing the curl in (111) in spherical polar coordinates gives

$$\mathbf{M} = \hat{\theta}\frac{1}{r\sin\theta}\frac{\partial(r\psi_M)}{\partial\phi} - \hat{\phi}\frac{1}{r}\frac{\partial(r\psi_M)}{\partial\phi} \tag{113}$$

If we then construct $\nabla\times\nabla\times\mathbf{M}$ by elementary vector calculus operations in spherical polar coordinates, we find that this **M** as in Eqs. (111) or (112) satisfies the vector Helmholtz equation in the form of Eq. (110) (see[30], p. 415 and[35], pp. 1764-6), i.e.,

$$\nabla\times\nabla\times\mathbf{M} = k^2\mathbf{M} \tag{114}$$

if $\psi_M$ satisfies the scalar Helmholtz equation in the form Eq. (109), i.e.,

$$\nabla^2\psi_M + k^2\psi_M = 0 \tag{115}$$

So, now we have found one possible kind of solution **M** as in Eqs.(111) or (112) for the "transverse" solution of the vector Helmholtz equation (105) or (110), a solution that is based on an underlying scalar function $\psi_M$ satisfying the scalar Helmholtz equation.

Now we can propose another solution for the "transverse" field, which we can write as

$$\mathbf{N} = \frac{1}{k}\nabla\times\mathbf{M} \tag{116}$$

(The $1/k$ factor is introduced for convenience to give **N** and **M** the same physical dimensions.) Then

$$\nabla\times\mathbf{N} = \frac{1}{k}\nabla\times\nabla\times\mathbf{M} = k\mathbf{M} \tag{117}$$

where we have used Eq. (114). Hence

$$\nabla\times\nabla\times\mathbf{N} = k\nabla\times\mathbf{M} = k^2\mathbf{N} \tag{118}$$

So we have shown that this **N** as defined in Eq. (116) is also a solution of the vector Helmholtz equation for "transverse" fields.

So, finally, we have three solutions to the vector Helmholtz equation (105); a "longitudinal" solution **L**, Eq. (103), and two "transverse" solutions **M**, Eq. (111), and **N**, Eq. (116). Note that all of these are based on scalar functions that satisfy a scalar Helmholtz equation. When we solve those scalar equations in spherical coordinates, we get the same radial behaviors as we found for scalar waves.

## S2.3 Explicit forms and sets of solutions for vector waves in spherical coordinates

Presuming for the moment that all these (vector) solutions **L**, **M**, and **N** can exist (the **L** solution will not exist for electromagnetic fields at finite frequencies, as we show later), we can now use known solutions of the scalar Helmholtz equation to construct sets of these vector solutions. In particular, in spherical polar coordinates in uniform isotropic media, we can propose for the scalar solutions $\phi_F$ or $\psi_M$ any of the functions (Eq. (84) above)

$$U_{nm}(\mathbf{r}) = z_n(kr) Y_{nm}(\theta,\phi) \equiv \frac{1}{kr} \zeta_n(kr) Y_{nm}(\theta,\phi)$$

We remember that $z_n(kr)$ are any of the spherical Bessel functions (or linear combinations of them) for a given order $n$, $\zeta_n$ are similarly any of the Riccati-Bessel functions, and $Y_{nm}$ are the spherical harmonics.

We can form the **L** solutions (if they exist) using one of the $U_{nm}(\mathbf{r})$ functions for $\phi_L$ in Eq. (103). For consistent physical dimensions, we introduce a constant factor $1/k$, giving

$$\begin{aligned}\mathbf{L}_{mn}(r,\theta,\phi) &= \frac{1}{k} \nabla U_{nm} = \frac{1}{k} \nabla \left[ z_n(kr) Y_{nm}(\theta,\phi) \right] \\ &= \hat{\mathbf{r}} \frac{1}{k} \frac{dz_n(kr)}{dr} Y_{nm}(\theta,\phi) + \frac{1}{k} z_n(kr) \nabla Y_{nm}(\theta,\phi) \\ &= \frac{dz_n(kr)}{d(kr)} \hat{\mathbf{r}} Y_{nm}(\theta,\phi) + \frac{z_n(kr)}{kr} r \nabla Y_{nm}(\theta,\phi)\end{aligned} \quad (119)$$

In terms of the vector spherical harmonics defined above, we can write this as

$$\mathbf{L}_{mn}(r,\theta,\phi) = \frac{dz_n(kr)}{d(kr)} \mathbf{P}_{mn}(\theta,\phi) + \frac{z_n(kr)}{kr} \mathbf{B}_{mn}(\theta,\phi) \quad n=0,1,2\ldots, \ -n \leq m \leq n \quad (120)$$

where the vector spherical harmonics $\mathbf{P}_{mn}$ and $\mathbf{B}_{mn}$ are given by Eqs. (92) and (93), respectively (and we note that $\mathbf{B}_{00}$ is the zero vector). Explicitly, in component form, we have, using Eqs. (92) and (95)

$$\begin{aligned}&\mathbf{L}_{mn}(r,\theta,\phi) = \\ &\left\{ \frac{dz_n(kr)}{d(kr)} P_n^m(\cos\theta) \hat{\mathbf{r}} + \frac{z_n(kr)}{kr} \left[ \frac{d}{d\theta} P_n^m(\cos\theta) \hat{\theta} + \frac{im}{\sin\theta} P_n^m(\cos\theta) \hat{\phi} \right] \right\} \exp(im\phi)\end{aligned} \quad (121)$$

Similarly, we construct

$$\begin{aligned}
\mathbf{M}_{mn}(r,\theta,\phi) &= \nabla\times[\mathbf{r}U_{nm}] = \nabla\times[\mathbf{r}z_n(kr)Y_{nm}(\theta,\phi)] \\
&= \nabla\times[z_n(kr)\{\mathbf{r}Y_{nm}(\theta,\phi)\}] \\
&= \nabla z_n(kr)\times\mathbf{r}Y_{nm}(\theta,\phi) + z_n(kr)\nabla\times[\mathbf{r}Y_{nm}(\theta,\phi)] \\
&= \hat{\mathbf{r}}\frac{dz_n(kr)}{dr}\times\mathbf{r}Y_{nm}(\theta,\phi) + z_n(kr)\mathbf{C}_{mn}(\theta,\phi)
\end{aligned} \qquad (122)$$

where we have used the vector calculus identity Eq. (96). Now using the fact that $\hat{\mathbf{r}}\times\mathbf{r}=0$ because these vectors are parallel, the first term on the right on the last line of Eq. (122) vanishes, leaving

$$\mathbf{M}_{mn}(r,\theta,\phi) = z_n(kr)\mathbf{C}_{mn}(\theta,\phi) \quad n=1,2,\ldots,\ -n\leq m\leq n \qquad (123)$$

where $\mathbf{C}_{mn}$ is given by Eq.(94). Explicitly in component form, using Eq. (99)

$$\mathbf{M}_{mn}(r,\theta,\phi) = z_n(kr)\left[\frac{im}{\sin\theta}P_n^m(\cos\theta)\hat{\theta} - \frac{d}{d\theta}P_n^m(\cos\theta)\hat{\phi}\right]\exp(im\phi) \qquad (124)$$

We can construct the set $\mathbf{N}_{mn}$ starting from the $\mathbf{M}_{mn}$ functions using the component form of Eq. (124). (We need the component form because we need to use a result that lies outside of vector calculus and its identities.) For our vector function $\mathbf{M}_{mn} = M_{mn\theta}\hat{\theta} + M_{mn\phi}\hat{\phi}$ above with no component in the $\hat{\mathbf{r}}$ direction, explicitly for $\mathbf{N}_{mn} \equiv N_{mnr}\hat{\mathbf{r}} + N_{mn\theta}\hat{\theta} + N_{mn\phi}\hat{\phi}$,

$$\begin{aligned}
\mathbf{N}_{mn} &= \frac{1}{k}\nabla\times\mathbf{M}_{mn} \\
&= \frac{1}{kr\sin\theta}\left[\frac{\partial}{\partial\theta}(M_{mn\phi}\sin\theta) - \frac{\partial M_{mn\theta}}{\partial\phi}\right]\hat{\mathbf{r}} - \frac{1}{kr}\frac{\partial}{\partial r}(rM_{mn\phi})\hat{\theta} + \frac{1}{kr}\frac{\partial}{\partial r}(rM_{mn\theta})\hat{\phi}
\end{aligned} \qquad (125)$$

For the $\hat{\theta}$ and $\hat{\phi}$ parts, we have

$$\begin{aligned}
N_{mn\theta}\hat{\theta} + N_{mn\phi}\hat{\phi} &= -\frac{1}{kr}\frac{\partial}{\partial r}(rM_{mn\phi})\hat{\theta} + \frac{1}{kr}\frac{\partial}{\partial r}(rM_{mn\theta})\hat{\phi} \\
&\equiv \left\{\frac{1}{kr}\frac{d[rz_n(kr)]}{dr}\frac{d}{d\theta}P_n^m(\cos\theta)\hat{\theta} + \frac{1}{kr}\frac{d[rz_n(kr)]}{dr}\frac{im}{\sin\theta}P_n^m(\cos\theta)\hat{\phi}\right\}\exp(im\phi) \\
&= \frac{1}{kr}\frac{d[rz_n(kr)]}{dr}\left\{\frac{d}{d\theta}P_n^m(\cos\theta)\hat{\theta} + \frac{im}{\sin\theta}P_n^m(\cos\theta)\hat{\phi}\right\}\exp(im\phi) \\
&= \frac{1}{kr}\frac{d[krz_n(kr)]}{d(kr)}\mathbf{B}_{mn}
\end{aligned} \qquad (126)$$

The $\hat{\mathbf{r}}$ component becomes

$$N_{nmr} = \frac{1}{kr\sin\theta}\left[\frac{\partial}{\partial\theta}(M_{mn\phi}\sin\theta) - \frac{\partial M_{mn\theta}}{\partial\phi}\right]$$
$$= \frac{1}{kr}z_n(kr)\left[-\frac{1}{\sin\theta}\frac{\partial}{\partial\theta}\left(\sin\theta\frac{dP_n^m(\cos\theta)}{d\theta}\right) + \frac{m^2}{\sin^2\theta}P_n^m\cos\theta\right]\exp(im\phi) \quad (127)$$

Changing variables from $\theta$ to $\eta = \cos\theta$, and noting that

$$\frac{d}{d\theta} = \frac{d\cos\theta}{d\theta}\frac{d}{d\cos\theta} = -\sin\theta\frac{d}{d\cos\theta} \quad (128)$$

and $\sin^2\theta = 1 - \cos^2\theta \equiv 1 - \eta^2$, we have

$$N_{nmr} = \frac{1}{kr}z_n(kr)\left\{\frac{\partial}{\partial\cos\theta}\left[-\sin^2\theta\frac{dP_n^m(\cos\theta)}{d\cos\theta}\right] + \frac{m^2}{\sin^2\theta}P_n^m(\cos\theta)\right\}\exp(im\phi)$$
$$= \frac{1}{kr}z_n(kr)\left\{\frac{m^2}{(1-\eta^2)}P_n^m(\eta) - \frac{d}{d\eta}\left[(1-\eta^2)\frac{dP_n^m(\eta)}{d\eta}\right]\right\}\exp(im\phi) \quad (129)$$

Now we recognize that we can substitute for the term in braces {} using the defining differential equation for the associated Legendre functions, Eq. (47), obtaining for the $\hat{\mathbf{r}}$ part of $\mathbf{N}_{mn}$

$$N_{nmr}\hat{\mathbf{r}} = n(n+1)\frac{1}{kr}z_n(kr)P_n^m(\cos\theta)\exp(im\phi)\hat{\mathbf{r}} \equiv n(n+1)\frac{1}{kr}z_n(kr)\mathbf{P}_{mn} \quad (130)$$

so finally we have, in terms of the vector spherical harmonics,

$$\mathbf{N}_{mn} = n(n+1)\frac{1}{kr}z_n(kr)\mathbf{P}_{mn} + \frac{1}{kr}\frac{d[krz_n(kr)]}{d(kr)}\mathbf{B}_{mn} \quad n = 1, 2..., -n \leq m \leq n \quad (131)$$

$\mathbf{N}_{00}$ is omitted (so $n$ starts from 1) because $\mathbf{B}_{00}$ is the zero vector as discussed in its definition above and, though $\mathbf{P}_{00}$ is not in general the zero vector, it is premultiplied by the factor $n$ in Eq. (131), so there is no contribution for $n = 0$.

To emphasize an important point from these results, while $\mathbf{L}_{mn}$ is defined (Eq. (120)) for all $n = 0, 1, 2, \ldots$, both $\mathbf{M}_{mn}$ and $\mathbf{N}_{mn}$ are defined (Eqs. (123) and (131)) starting from $n = 1$, i.e., for $n = 1, 2, \ldots$.

(In our constructions above of $\mathbf{L}_{mn}$, $\mathbf{M}_{mn}$, and $\mathbf{N}_{mn}$, we have avoided normalizing factors for simplicity since we do not need them, but normalized versions are given in [33], p. 27.)

### S2.4 Near and far field vector waves in spherical coordinates

We can conveniently divide the various terms in $\mathbf{L}_{mn}$, $\mathbf{M}_{mn}$, and $\mathbf{N}_{mn}$ into what we can call "near-field" and "far-field" terms. Here by "near-field" we mean terms that certainly fall off faster than $1/r$ at larger radii $r$, so they cannot correspond to waves that propagate in the far field. By "far-field", conversely, we mean terms that fall off with an underlying $1/r$ dependence, so they can correspond to propagating spherical waves of constant power for large $r$.

We note that the various spherical Bessel functions $z_n(\rho)$ (i.e., explicitly, $j_n(\rho)$, $y_n(\rho)$, $h_n^{(1)}(\rho)$ or $h_n^{(2)}(\rho)$) all have an underlying $1/r$ fall-off in their magnitude at large $r$ (as well as a quasi-oscillatory behavior with $r$ at large enough $r$). So, for example, a term in $z_n(kr)/kr$ is falling off as $\sim 1/r^2$; so, it is a "near-field" term in our categorization. A term with the simple derivative $dz_n(kr)/dkr$, however, because of the quasi-oscillatory part, has a large component that falls off as $\sim 1/r$, so we categorize it as a "far-field" term here. (Note explicitly that $dz_n(\rho)/d\rho = (n/\rho)z_n(\rho) - z_{n+1}(\rho)$, in which the component in $z_{n+1}(\rho)$ falls off as $\sim 1/\rho$). A function $krz_n(kr)$ is essentially independent of $r$ (in its overall magnitude) at large $r$. (In fact, it corresponds also to a Riccati-Bessel function, which we have explicitly seen above tends to a constant magnitude of quasi-oscillation at large $r$.) A term with the derivative $d(krz_n(kr))/dkr$ is therefore (in magnitude of quasi-oscillation) also essentially independent of $r$ at large $r$; so a term in $(1/kr)d(krz_n(kr))/dkr$ is falling of as $\sim 1/r$ at large $r$, so it is a "far-field" term.

With these categorizations of near- and far-field terms indicated by superscripts, we can write for $\mathbf{L}_{mn}$

$$\mathbf{L}_{mn} = \mathbf{L}_{mn}^{(near)} + \mathbf{L}_{mn}^{(far)} \tag{132}$$

from Eq. (120) with

$$\mathbf{L}_{mn}^{(near)} = \frac{z_n(kr)}{kr}\mathbf{B}_{mn}(\theta,\phi) \tag{133}$$

and for $\mathbf{N}_{mn}$

$$\mathbf{N}_{mn} = \mathbf{N}_{mn}^{(near)} + \mathbf{N}_{mn}^{(far)} \tag{134}$$

from Eq. (131) with

$$\mathbf{N}_{mn}^{(near)} = n(n+1)\frac{1}{kr}z_n(kr)\mathbf{P}_{mn} \tag{135}.$$

$\mathbf{M}_{mn}$ does not have any corresponding near-field terms, so we can collect and state all the far-field terms as

$$\mathbf{L}_{mn}^{(far)} = \frac{dz_n(kr)}{d(kr)}\mathbf{P}_{mn}(\theta,\phi) \quad n = 0,1,2..., \ -n \leq m \leq n \tag{136}$$

$$\mathbf{M}_{mn}(r,\theta,\phi) = z_n(kr)\mathbf{C}_{mn}(\theta,\phi) \quad n = 1,2..., \ -n \leq m \leq n \tag{137}$$

$$\mathbf{N}_{mn}^{(far)} = \frac{1}{kr}\frac{d[krz_n(kr)]}{d(kr)}\mathbf{B}_{mn}(\theta,\phi) \quad n = 1,2..., \ -n \leq m \leq n \tag{138}$$

(Eq. (137) is the same as Eq. (123) above.) The allowed ranges for *n* (including in particular whether $n = 0$ is allowed) follow from the corresponding ranges for $\mathbf{L}_{mn}$, $\mathbf{M}_{mn}$, and $\mathbf{N}_{mn}$ in Eqs. (92), (93), and (94).

We see also that these far-field terms have simple polarization behavior. $\mathbf{L}_{mn}$ is radially polarized, and $\mathbf{M}_{mn}$ and $\mathbf{N}_{mn}^{(far)}$ have transverse polarizations that are also orthogonal to one another. Remember Eq. (101), $\mathbf{B}_{mn} \cdot \mathbf{C}_{mn} = 0$, so

$$\mathbf{M}_{mn} \cdot \mathbf{N}_{mn}^{(far)} = 0 \tag{139}$$

(It is also quite generally true that $\mathbf{M}_{mn}$ (as in Eq. (123)) and $\mathbf{N}_{mn}$ (as in Eq. (131)) have orthogonal polarizations because $\mathbf{P}_{nm}$, being radially polarized, is orthogonal to the polarizations of both $\mathbf{B}_{mn}$ and $\mathbf{C}_{mn}$.)

## S2.5 Electromagnetic waves in spherical coordinates

So far, we have considered vector fields quite generally that satisfy a vector Helmholtz equation in spherical coordinates, leading to sets of solutions $\mathbf{L}_{mn}$, $\mathbf{M}_{mn}$, and $\mathbf{N}_{mn}$ that are based on vector spherical harmonics $\mathbf{P}_{mn}$, $\mathbf{B}_{mn}$, and $\mathbf{C}_{mn}$ and spherical Bessel functions (or, equivalently, Riccati-Bessel functions). Sound waves in air and some kinds of elastic waves in solids can satisfy such an equation (or a slightly extended version with different transverse and longitudinal wave velocities, as appropriate for a homogeneous isotropic solid material, for example[35]). Indeed, sounds waves and such elastic materials are known to support the kinds of longitudinally polarized waves (e.g., velocity waves for sound) that can be described using the $\mathbf{L}_{mn}$ functions. Though we will not consider such sound or elastic waves further, we could use exactly the techniques in this work to analyze the numbers of such propagating waves and their tunneling escape.

The vector waves of most interest to us will be electromagnetic waves. As we demonstrate below, those satisfy the vector Helmholtz equation, though only the $\mathbf{M}_{mn}$ and $\mathbf{N}_{mn}$ sets of functions are required to describe them in spherical coordinates. The $\mathbf{L}_{mn}$ waves are not present, as we will show.

We start with Maxwell's equations for the electric and magnetic fields $\mathbf{E}$ and $\mathbf{B}$, respectively, for the case of a uniform isotropic material, so with constant permittivity $\varepsilon$ and permeability $\mu$, and with no net charge density or current density.

$$\nabla \cdot \mathbf{E} = 0 \tag{140}$$

$$\nabla \cdot \mathbf{B} = 0 \tag{141}$$

$$\nabla \times \mathbf{E} = -\frac{\partial \mathbf{B}}{\partial t} \tag{142}$$

$$\nabla \times \mathbf{B} = \mu\varepsilon \frac{\partial \mathbf{E}}{\partial t} \tag{143}$$

So

$$\nabla \times \nabla \times \mathbf{E} = \nabla \times \left(-\frac{\partial \mathbf{B}}{\partial t}\right) = -\frac{\partial}{\partial t}(\nabla \times \mathbf{B}) = -\mu\varepsilon \frac{\partial^2 \mathbf{E}}{\partial t^2}$$

So

$$\nabla \times \nabla \times \mathbf{E} + \mu\varepsilon \frac{\partial^2 \mathbf{E}}{\partial t^2} = 0 \tag{144}$$

Similarly

$$\nabla \times \nabla \times \mathbf{B} = \mu\varepsilon \nabla \times \left(\frac{\partial \mathbf{E}}{\partial t}\right) = \mu\varepsilon \frac{\partial}{\partial t}(\nabla \times \mathbf{E}) = -\mu\varepsilon \frac{\partial^2 \mathbf{B}}{\partial t^2}$$

So

$$\nabla \times \nabla \times \mathbf{B} + \mu\varepsilon \frac{\partial^2 \mathbf{B}}{\partial t^2} = 0 \tag{145}$$

As usual, we can identify

$$\mu\varepsilon = 1/v^2 \tag{146}$$

where $v$ is the wave velocity. Since we are only interested in monochromatic fields at some (angular) frequency $\omega$, so with a presumed time dependence $\propto \exp(-i\omega t)$, Eqs. (144) and (145) become, respectively

$$\nabla \times \nabla \times \mathbf{E} = k^2 \mathbf{E} \tag{147}$$

$$\nabla \times \nabla \times \mathbf{B} = k^2 \mathbf{B} \tag{148}$$

where the wavevector magnitude $k$ is given by

$$k = \frac{\omega}{v} = \sqrt{\mu\varepsilon}\,\omega$$

Noting the vector calculus identity $\nabla^2 \mathbf{f} \equiv \nabla(\nabla \cdot \mathbf{f}) - \nabla \times \nabla \times \mathbf{f}$, Eq. (106), and noting that both $\nabla \cdot \mathbf{E}$ and $\nabla \cdot \mathbf{B} = 0$ (Eqs. (140) and (141)), we can if we wish rewrite these as the full vector Helmholtz equations

$$\nabla^2 \mathbf{E} + k^2 \mathbf{E} = 0 \tag{149}$$

$$\nabla^2 \mathbf{B} + k^2 \mathbf{B} = 0 \tag{150}$$

Though, in fact, Eqs. (148) and (147) are all we need.

Suppose now we consider the $\mathbf{E}$ field in the full Helmholtz decomposition as in Eq. (102), which would lead to a representation

$$\mathbf{E} = \nabla \phi_E + \nabla \times \mathbf{A}_E \tag{151}$$

for some formal electric scalar and vector potentials $\phi_E$ and $\mathbf{A}_E$. (Note, incidentally, that the $\phi_E$ defined this way would correspond to minus the electrostatic potential $\phi$ as conventionally defined in electrostatics, where for static fields $\mathbf{E} = -\nabla\phi$.) However, from the Maxwell equation (140) in the absence of net charge density, $\nabla \cdot \mathbf{E} = 0$, and since generally $\nabla \cdot (\nabla \times \mathbf{f}) = 0$, we have

$$\nabla \cdot \mathbf{E} = \nabla^2 \phi_E = 0 \qquad (152)$$

(where we note $\nabla \cdot \nabla f \equiv \nabla^2 f$). But if the "longitudinal" component $\mathbf{L} = \nabla \phi_E$ is to be a solution of the vector Helmholtz equation, we require that $\phi_E$ satisfies the scalar Helmholtz equation (109), i.e., $\nabla^2 \phi_E + k^2 \phi_E = 0$. Since $\nabla^2 \phi_E = 0$ (Eq. (152)), we can only have a finite $\mathbf{L} = \nabla \phi_E$ for $k = 0$, which corresponds to a static electric field (which satisfies Laplace's equation for no charge density, $\nabla^2 \phi_E = 0$). So, for finite frequencies, there is no longitudinal field component $\mathbf{L}$ for the electric field.

Since also $\nabla \cdot \mathbf{B} = 0$, Eq. (141), we have an identical argument to discard the longitudinal field component $\mathbf{L}$ for the magnetic field. Hence, for both electric and magnetic fields at a finite frequency, we need only the $\mathbf{M}_{mn}$ and $\mathbf{N}_{mn}$ functions to describe them (each of which satisfies the equations (147) and (148)).

Hence, in summary conclusion, electric and magnetic fields in uniform isotropic media with no charge density or current density, at some frequency $\omega$ with corresponding wavevector magnitude $k$, can be represented in spherical coordinates using the two sets of functions $\mathbf{M}_{mn}(r,\theta,\phi)$, Eq. (123), and $\mathbf{N}_{mn}(r,\theta,\phi)$, Eq. (131), and in simpler form in the far field by $\mathbf{M}_{mn}(r,\theta,\phi)$ (as before) and $\mathbf{N}_{mn}^{(far)}(r,\theta,\phi)$, Eq. (138). In all cases, $n = 1, 2..., -n \leq m \leq n$. Note again that there are no $n = 0$ solutions for these electromagnetic waves.

So, what we have shown for electromagnetic waves in uniform isotropic media is that

- (i) There is no $\mathbf{L}$ spherical wave.
- (ii) For each of the electric field $\mathbf{E}$ and the magnetic field $\mathbf{B}$, there are both $\mathbf{M}$ and $\mathbf{N}$ spherical waves.
- (iii) There are no $n = 0$ spherical waves.

All that remains now is to write out these results for the $\mathbf{M}$ and $\mathbf{N}$ waves for the electric and magnetic fields in some conventional form.

## S2.6 Relation to multipole expansions

The electromagnetic results we have derived above correspond conventionally to the multipole representation of electromagnetic waves. Jackson[25], pp. 430-1, defines two sets of fields, in terms of electric field $\mathbf{E}$ and magnetic field $\mathbf{H} = \mathbf{B}/\mu$. Historically, these multipole fields are sometimes known as magnetic and electric multipoles, though that notation can be confusing. Instead, we use the alternate notation of "transverse electric" (TE) and "transverse magnetic" (TM) multipoles, which relates more naturally to propagating wave behavior. In our notations for $n$, $m$, and with $z_n(kr)$ as any linear combination of spherical Bessel functions of order $n$, we can write these results as follows.

(i) The transverse electric (TE) (or "magnetic") multipole of order ($n$, $m$), is given by

$$\mathbf{E}_{nm}^{(TE)}(r,\theta,\phi) = -i\sqrt{\frac{\mu}{\varepsilon}}\, z_n(kr)(\mathbf{r}\times\nabla)Y_{nm}(\theta,\phi) \tag{153}$$

$$\mathbf{H}_{nm}^{(TE)}(r,\theta,\phi) = -i\frac{1}{k}\sqrt{\frac{\varepsilon}{\mu}}\,\nabla\times\mathbf{E}_{nm}^{(TE)} \tag{154}$$

Using Eqs. (98) and (123), Eq. (153) becomes

$$\mathbf{E}_{nm}^{(TE)}(r,\theta,\phi) = i\sqrt{\frac{\mu}{\varepsilon}}\, z_n(kr)\mathbf{C}_{mn}(\theta,\phi) \equiv i\sqrt{\frac{\mu}{\varepsilon}}\,\mathbf{M}_{mn}(r,\theta,\phi) \tag{155}$$

Using Eq. (116), Eq. (154) becomes

$$\mathbf{H}_{nm}^{(TE)}(r,\theta,\phi) = \mathbf{N}_{mn}(r,\theta,\phi) \tag{156}$$

(ii) The transverse magnetic (TM) (or "electric") multipole of order $(n, m)$ is given by

$$\mathbf{H}_{nm}^{(TM)}(r,\theta,\phi) = -i\, z_n(r)(\mathbf{r}\times\nabla)Y_{nm}(\theta,\phi) \tag{157}$$

$$\mathbf{E}_{nm}^{(TM)}(r,\theta,\phi) = i\frac{1}{k}\sqrt{\frac{\mu}{\varepsilon}}\,\nabla\times\mathbf{H}_{nm}^{(TM)} \tag{158}$$

Using Eqs. (98) and (123), Eq. (157) becomes

$$\mathbf{H}_{nm}^{(TM)}(r,\theta,\phi) = i\, z_n(kr)\mathbf{C}_{mn}(\theta,\phi) \equiv i\mathbf{M}_{mn}(r,\theta,\phi) \tag{159}$$

and using Eq. (116), Eq. (158) becomes

$$\mathbf{E}_{nm}^{(TM)}(r,\theta,\phi) = -\sqrt{\frac{\mu}{\varepsilon}}\,\mathbf{N}_{nm}(r,\theta,\phi) \tag{160}$$

Hence, the $\mathbf{M}_{mn}$ and $\mathbf{N}_{mn}$ functions are basis functions for these expansions. So, our representation of electromagnetic fields in terms of $\mathbf{M}_{mn}(r,\theta,\phi)$ and $\mathbf{N}_{mn}(r,\theta,\phi)$ is the same as the multipole expansion of electromagnetic fields. Note still, as mentioned explicitly also by Jackson[25], p. 431, these functions are defined only for $n$ starting at 1, not zero.

For completeness, we can mention that, to obtain normalized versions of these multipole radiation functions for convenient expansions, Jackson [25], p. 431, introduces a set of functions

$$\mathbf{X}_{nm}(\theta,\phi) = \frac{-i}{\sqrt{n(n+1)}}(\mathbf{r}\times\nabla)Y_{nm}(\theta,\phi) \equiv \frac{i}{\sqrt{n(n+1)}}\mathbf{C}_{mn}(\theta,\phi) \tag{161}$$

in terms of which

$$\mathbf{M}_{mn}(r,\theta,\phi) = -i\sqrt{n(n+1)}\, z_n(kr)\mathbf{X}_{nm}(\theta,\phi) \tag{162}$$

$$\mathbf{N}_{mn} = \frac{1}{k}\nabla\times\mathbf{M}_{mn} = \frac{-i}{k}\sqrt{n(n+1)}\, z_n(kr)\nabla\times\mathbf{X}_{nm}(\theta,\phi) \tag{163}$$

These sets of functions $\mathbf{X}_{nm}$ are also explicitly orthogonal for different $n$ and/or different $m$.

Note, incidentally, that these definitions from Jackson[25] are not formally dimensionally correct. They get the relative dimensions of electric and magnetic fields correct, but they formally correspond to a dimensionless **H** field. To be dimensionally correct, all these fields would need to be multiplied by a constant with dimensions of amperes/meter (those being the units of the **H** field). (One possible reason for Jackson's choice is that there is no obvious combination of the fundamental constants $\varepsilon_o$ and $\mu_o$ and the velocity of light in electromagnetism itself that has the dimensions of either the **E** field or the **H** field.)

Incidentally, for a spherical shell volume enclosing some sources, the radius of that sphere is essentially telling us when we can stop including terms in a multipole expansion. Once the escape radius associated with a given *n* becomes sufficiently much larger than the sphere radius, we may stop including further terms in the expansion.

## S3 Communication modes for scalar spherical waves

The primary interest in this paper is how waves behave as we propagate out from a spherical surface. We have shown the various scalar (and vector) waves in spherical coordinates that could be considered as a basis for any such outgoing waves. There is an underlying way of thinking about waves that clarifies that these are a complete basis for any outgoing waves, and that explicitly links to the orthogonal channels for communication in such systems. This is the communication mode approach[12], which establishes the orthogonal source functions in a source space that communicate to resulting orthogonal waves in a receiving space.

This approach is based on the singular-value decomposition of the coupling operator $\mathsf{G}_{SR}$ (or equivalently, the Green's function $G_{SR}(\mathbf{r}_R, \mathbf{r}_S)$) between the source and receiver spaces. It rigorously and uniquely (within geometric degeneracies) gives these pairs of basis functions in the source and receiver spaces. The corresponding waves in the space between the source and receiver spaces can be calculated from these, but it is important to understand that the eigenfunctions and orthogonalities here are for the functions in the source space and the receiving space.

If we consider, for example, a spherical or spherical shell source space, of (outer) radius $r_{So}$ and a (concentric) spherical shell receiving space with some larger (inner) radius $r_{Ro}$ (so there is space between these two spherical surfaces) as in Fig. S13, we can rigorously and analytically calculate these orthogonal source and received wave functions in those spaces. We take thicknesses $\Delta r_{So}$ and $\Delta r_{Ro}$, respectively, for these source and receiver shells that we presume to be very small, both compared to a wavelength and compared to the radii of the shells.

As we construct these volumes and the associated communication modes, importantly, we will see that the resulting waves we can calculate for the space between these spherical surfaces correspond exactly to the spherical waves we have established above. So, these spherical waves correspond to the waves associated with the communication modes of such a spherical system. This guarantees that these waves are complete for describing any such outgoing wave because we know from the communication mode analysis that the waves in the spherical shell receiving volume form a complete set for describing such outgoing waves.

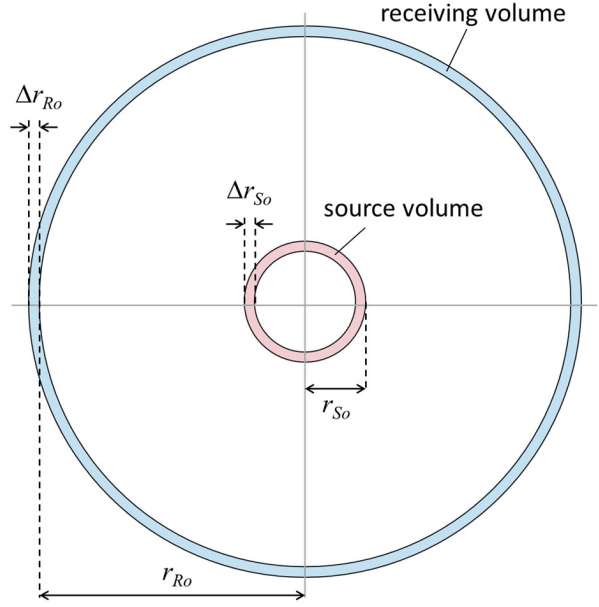

Fig. S13. Illustration of thin, concentric spherical shell source and receiving volumes.

Note that, as is generally the case for communication modes, the waves in the space between the source and receiving volumes are *not* the eigensolutions of the problem – they are not the "modes". In fact, in the present problem, though each one of the waves in that space can be described as the solution of a differential equation that is in eigen equation form, Eqs. (7) (in the main text), (76) or (77), the equation is different for different $n$; explicitly, the potential energy is different for every $n$. These waves for different $n$ are *not* solutions of the same differential equation; they are *not* different eigensolutions of the same equation.

In contrast, the communication modes are sets of functions, one for the source space and one for the receiving space, that are each sets solutions of the same eigenequation (with a different, though related, eigenequation for each of the two spaces, source and receiving). These sets of functions in each space, because they are the eigenfunctions of compact Hermitian operators in each case, do form orthogonal and complete sets of functions for describing any source in the source space and any corresponding resulting waves in the receiving space. To repeat, the waves we have calculated above turn out to be the waves associated with the communication modes between source and receiver spaces with concentric spherical surfaces, and this proves the completeness of these waves for describing any outgoing (scalar) wave from a spherical surface.

Now let us formally solve for the communication modes for this spherical shell problem. Quite generally, we can write the coupling operator in the singular-value decomposition (SVD) form[12]

$$\mathsf{G}_{SR} \equiv \sum_j \sigma_j |\phi_{Rj}\rangle\langle\psi_{Sj}| \qquad (164)$$

Here the Dirac "ket" notation $|\phi_{Rj}\rangle$ is a function representing a wave in the receiving space, and is one of the complete set of functions (indexed by $j$) in that space that results from solving the eigen equation $\mathsf{G}_{SR}\mathsf{G}_{SR}^\dagger |\phi_{Rj}\rangle = |\sigma_j|^2 |\phi_{Rj}\rangle$ where the "†" indicates the Hermitian adjoint (conjugate transpose) of the operator (or matrix). Similarly, $|\psi_{Sj}\rangle$ is one of the complete set of eigenfunctions of the equation $G_{SR}^\dagger G_{SR} |\psi_{S_j}\rangle = |\sigma_j|^2 |\psi_{S_j}\rangle$. (The "bra" $\langle\psi_{Sj}|$ is the Hermitian adjoint of $|\psi_{Sj}\rangle$; if we think of these as mathematical vectors, the "ket" is a column vector, and its "bra" version is a row vector with complex conjugated elements.) Note $G_{SR} G_{SR}^\dagger$ and $G_{SR}^\dagger G_{SR}$ are both Hermitian operators and have provably non-negative eigenvalues[12].

Conventionally, we choose these eigenfunctions to be normalized; together with their known orthogonality (which follows from them being eigenfunctions of Hermitian operators), we can write the "orthonormality" relations $\langle\phi_{Rp}|\phi_{Rq}\rangle = \delta_{pq}$ and $\langle\psi_{Sp}|\psi_{Sq}\rangle = \delta_{pq}$, where $\delta_{pq}$ is the Kronecker delta, with $\delta_{pq} = 0, p \ne q; \delta_{pq} = 1, p = q$. Because of this orthonormality, we can see from Eq. (164) that

$$G_{SR} |\psi_{Sj}\rangle = \sigma_j |\phi_{Sj}\rangle \qquad (165)$$

That is, each one of the orthogonal source functions $|\psi_{Sj}\rangle$ leads to the generation of the corresponding one $|\phi_{Rj}\rangle$ of the orthogonal received wave functions, with an (amplitude) coupling strength given by $\sigma_j$, which is known as the singular value, and which can be a complex number, corresponding to the specific phase relationship between the source and received functions.

Hence the SVD gives a set of orthogonal source functions that couple one by one to a set of orthogonal resulting waves, and this can be viewed as the physical meaning of the SVD in this case. Note that this decomposition Eq. (164) is unique (other than for symmetry degeneracies as usual with eigen problems). So, if we find a set of orthogonal functions in the two spaces that represent the operator in this way, we have found the SVD of the operator.

Now, the free-space Green's function of the Helmholtz equation (1) can be written (Jackson[25], Eq. 9.98 on p. 428, also Hanson and Yakovlev[38], Eq. 5.172, Varshalovich et al.[39], p. 165, Eq. (19))

$$G(\mathbf{r},\mathbf{r}') = \frac{\exp(ik|\mathbf{r}-\mathbf{r}'|)}{4\pi|\mathbf{r}-\mathbf{r}'|} = ik \sum_{n=0}^{\infty} \sum_{m=-n}^{n} j_n(kr') Y_{nm}^*(\theta',\phi') h_n^{(1)}(kr) Y_{nm}(\theta,\phi) \qquad (166)$$

where we can regard $\mathbf{r}'$ with spherical coordinates $(r',\theta',\phi')$ as being the position of the source and $\mathbf{r}$ with coordinates $(r,\theta,\phi)$ as being the position at which we examine the resulting wave. Note immediately that, because the spherical harmonic functions are complete and orthogonal for describing functions of angle in spherical coordinates, this expansion is already close to the form of Eq. (164). If our source space is a thin spherical shell, the functions $j_n(kr')Y_{nm}^*(\theta',\phi')$ are already an orthogonal set in this space because the spherical harmonics for different $n$ and/or $m$ are orthogonal; we do not even need any orthogonality properties from the spherical Bessel functions $j_n$. Similarly, for a receiving space that is a thin spherical shell, the functions $h_n^{(1)}(kr)Y_{nm}(\theta,\phi)$ are also orthogonal for different $n$ and/or $m$.

To complete the expression as in Eq. (164), we need to express the Green's function in terms of orthogonal functions that are also normalized. Formally, then, in the source space, we want to choose the constant $A_{Sn}$ such that

$$k \int_{V_S} |A_{Sn}|^2 \, j_n(kr')^2 \, |Y_{nm}(\theta',\phi')|^2 \, r'^2 \sin\theta' dr' d\theta' d\phi' = 1 \tag{167}$$

where $V_S$ is the volume of the source spherical shell as in Fig. S13. (Note that $j_n$ is real, so we do not need to take its modulus squared in this integral. Note also that we will split the "$k$" prefactor between the two functions when normalizing, assigning a factor $\sqrt{k}$ to both functions.)

The integral over the angles is already given by the orthonormality relation for the spherical harmonics, Eq. (58) (with the resulting value of 1). Because the spherical shell is presumed very thin, we can remove the $j_n(kr')^2$ term to outside the integral as $j_n(kr_{So})^2$, leaving

$$|A_{Sn}|^2 \, j_n(kr_{So})^2 \, k \int_{r_{So}-\Delta r_S}^{r_{So}} r'^2 dr' \simeq |A_{Sn}|^2 \, j_n(kr_{So})^2 \, r_{So}^2 k \Delta r_S = 1 \tag{168}$$

So, we can choose a normalization factor

$$A_{Sn} = 1/\left[ j_n(kr_{So}) r_{So} \sqrt{k\Delta r_S} \right] \tag{169}$$

The source basis functions become, explicitly,

$$|\psi_{Snm}\rangle \equiv \frac{1}{j_n(kr_{So}) r_{So} \sqrt{k\Delta r_S}} j_n(kr_{So}) Y_{nm}(\theta',\phi') = \frac{1}{r_{So} \sqrt{k\Delta r_S}} Y_{nm}(\theta',\phi') \tag{170}$$

where we have replaced $j_n(kr')$ by $j_n(kr_{So})$ given our thin spherical shell volume. Similarly, we can choose the normalized receiving function as

$$|\phi_{Rnm}\rangle \equiv \frac{1}{r_{Ro} \sqrt{k\Delta r_R}} Y_{nm}(\theta,\phi) \tag{171}$$

$$|\phi_{Rnm}\rangle \equiv \frac{1}{r_{Ro} \sqrt{k\Delta r_R}} Y_{nm}(\theta,\phi) \tag{172}$$

Using these normalized forms, we can rewrite Eq. (166) in the form

$$G(\mathbf{r},\mathbf{r}') \equiv G_{SR} \equiv \sum_{n=0}^{\infty} \sum_{m=-n}^{n} \sigma_{nm} |\phi_{Rnm}\rangle\langle\psi_{Snm}| \tag{173}$$

where

$$\sigma_{nm} = ikr_{So}r_{Ro}\sqrt{\Delta r_S \Delta r_R} \, j_n(kr_{So}) h_n^{(1)}(kr_{Ro}) \tag{174}$$

If we are interested in propagation to the far field, so for very large $r_{Ro}$, we can use the far-field approximation to the spherical Hankel function, Eq. (70), and write

$$|\sigma_{nm}|^2 \simeq r_{So}^2 \, \Delta r_S \Delta r_R \left[ j_n(kr_{So}) \right]^2 \tag{175}$$

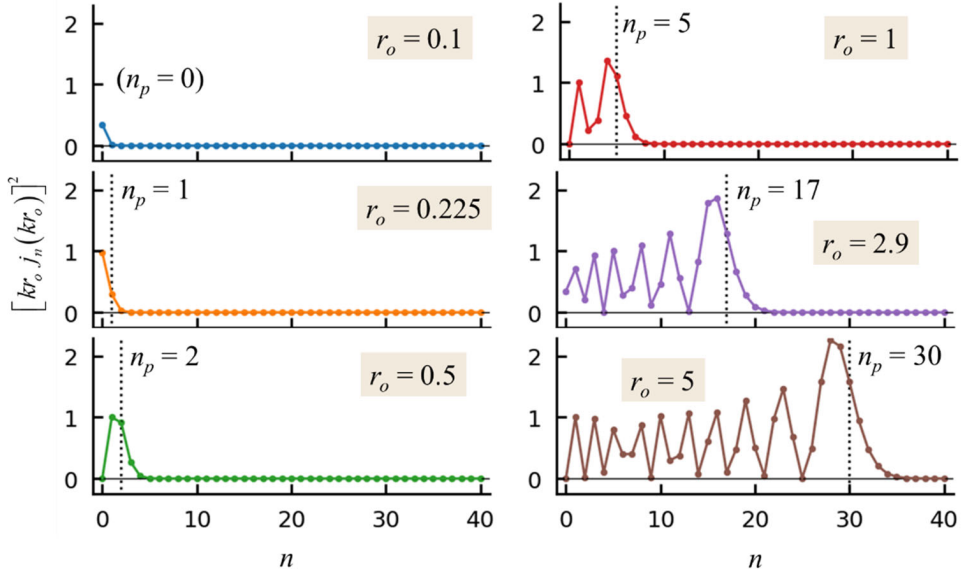

Fig. S14. Plot of $[kr_o j_n(kr_o)]^2$, a dimensionless version of the modulus squared of the singular values, for the communication modes between sources in a thin spherical shell of radius $r_o$ wavelength and another thin spherical shell of very large radius, against the parameter $n$ of the spherical waves. (Lines are to guide the eye; only the points are meaningful.) These plots are for a representative set of values of $r_o$ – specifically, 0.1, 0.225, 1, 2.9, and 5 wavelengths. The corresponding values for $n_p$, the largest value of $n$ for which all waves start out as propagating, are also shown.

We have plotted a related dimensionless version of these squared singular values, $[kr_o j_n(kr_o)]^2$, in Fig. S14 for various values $r_o$ of $r_{So}$. For $n > n_p$, in each case these squared singular values fall off rapidly and monotonically with increasing $n$, which corresponds to waves that start out by tunneling. For smaller values of $n$, there is a somewhat oscillatory behavior in the singular value magnitudes, which can be rationalized as the result of interference between sources at different points on the thin spherical surface. The vector electromagnetic waves follow a similar communication mode analysis that we have presented in[14].

For our present discussion, the specific results for the singular values and/or its square are less important than the fact that we have shown that we have the singular value decomposition of the operator connecting our thin spherical shell source and receiver volumes, and we see that the wave functions consistent with the communication modes as a function of distance outside the source spherical shell are exactly of the form $h_n^{(1)}(kr) Y_{nm}(\theta, \phi)$.

## S4 Inward waves

As is common in many wave and antenna problems, we have examined the wave propagation with a "radiation condition", that is, we have examined only outward propagating waves so far. This is the solution when there are sources within our spherical volume, and none outside. We can also consider situations where there is instead an incoming wave.

One common approach is to view this as a scattering problem, and many such investigations exist, especially with incident plane waves, such as Mie scattering (see, e.g.,[22] for a recent discussion), which does, incidentally use the same kind of spherical wave basis as we use here. Generally, an incident wave can give rise to some absorption in an object, which can be viewed as a source in the object that is excited by the incoming wave in such a way that it absorbs some of the incoming radiation, and may also radiate some new radiation of its own.

We will not investigate such scattering problems here. We can, however, briefly consider the solutions with our spherical waves in the situation where there is an incident wave on our spherical volume, but there are no sources inside the spherical volume. Then there can be no $y_n$ spherical Bessel function because the associated wave would become singular at $r = 0$. (For this reason, we also discard the corresponding $C_n$ Riccati-Bessel function in the solutions.) In such a case, the only corresponding radial solutions are the $j_n$ spherical Bessel functions, with the corresponding $S_n$ Riccati-Bessel function. These solutions are standing waves, not propagating waves; all waves that go "in" must come back "out" again.

For an $n = 0$ scalar wave, the solution simply corresponds to us focusing towards the center of the spherical volume, which would could view as a wave essentially passing through that focus, though note that there is no such $n = 0$ electromagnetic wave, as discussed above. However, as we consider waves with higher $n$, we start now to see the tunneling behavior emerging, this time as a tunneling penetration *into* the volume from the outside. See, for example, the $j_n$ spherical Bessel functions and the $S_n$ Riccati-Bessel functions in Fig. S4, which are the solid (blue) lines in each case. Once $n$ becomes larger than about 5, we can very clearly see the tunneling decay *into* the spherical surface below the escape radius.

In fact, such high-$n$ waves effectively never get far into the interior of the spherical surface, being essentially reflected off it, which we can view as forming the standing wave. See Fig. S15, which shows, in this example for $n = 17$, that the waves essentially do not penetrate into an inner spherical volume of about 2 wavelengths in radius. Note, then, that such waves cannot effectively "see" far inside the sphere. They are, in a sense, reflecting off empty space. Generally, the spherical volume of radius significantly less than the escape radius for a given $n$ is effectively "cloaked"[29] as far as those waves are concerned. Fig. S16 shows similar examples up to $n = 100$.

Note that such standing waves with total reflection from a spherically shaped region are part of the normal propagation of waves through free space. Indeed, to us, it would appear that the inward spherical wave is just propagating through free space to become the outward wave, as if it were focusing through the center of the sphere. But, in fact, for all waves of any substantial $n$ value, they are instead reflecting back on us, never getting to the center.

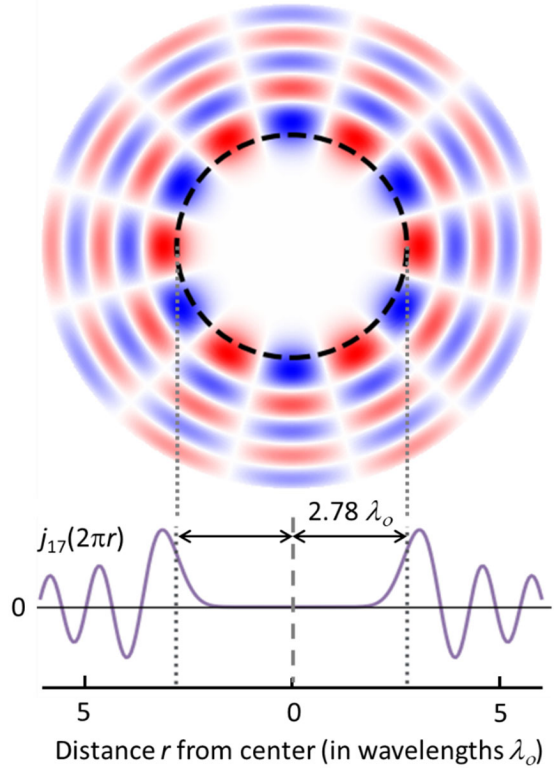

Fig. S15. Plot of a spherical standing wave, shown in the equatorial plane in the 2D color map plot (upper part of figure). This is plotted for $n=17$, $m=6$, for radii up to 6.05 wavelengths. For $n=17$, the escape radius is $\simeq 2.78$ wavelengths, shown also as the dashed circle. The corresponding radial spherical Bessel function $j_{17}(kr)$ (which is equivalent to $j_{17}(2\pi r)$ for $r$ in units of wavelengths) is shown in the lower graph.

Fig. S16 shows for different $n$ that the waves "starting" from the outside indeed initially become stronger as the radial distance decreases, as we would expect for spherical waves focusing toward the middle. For all $n$ larger than about 3, however, once they pass the escape radius, they rapidly get weak. They are failing to penetrate far inside the escape radius; they are instead effectively reflecting off this region of empty space.

Suppose we have some object located around the center of these plots in Figs. S15 and S16. For small $n$, we would have to deal with possible absorption and/or scattering for incoming waves because the wave clearly reaches into the center of the plots. (In such a case, we would have to solve the scattering problem properly.) With increasing $n$ in our incoming waves, once the sphere of the corresponding escape radius $r_{escn}$ becomes even moderately larger than the object, these waves of higher $n$ will effectively not see the object because they would not reach it. So, in microscopy, for example, if we were to try to illuminate the object with such high $n$ waves to expose some fine detail, if the object is much smaller than the corresponding escape radius, such probing waves would effectively not even reach the object. So, we see that, even in the case where we have some object that could scatter and/or absorb radiation, the number of external waves that we need to consider in interacting with that object is essentially the same as the number $N_p$ of waves that could escape from the bounding sphere round the object.

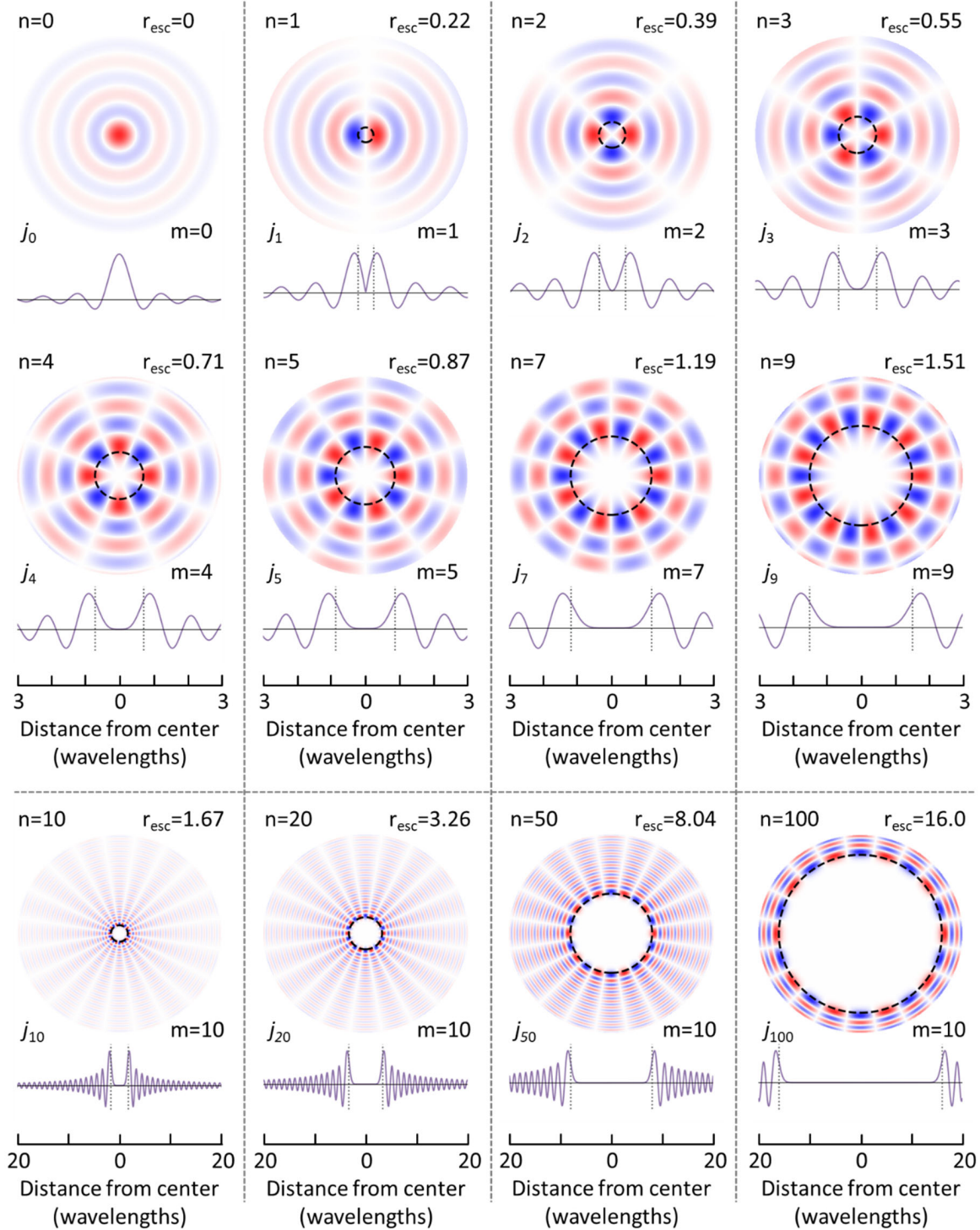

Fig. S16. Plots of spherical standing waves as in Fig. S15, for various values of $n$, together with the corresponding spherical Bessel functions $j_n$ and escape radii $r_{escn}$ in wavelengths. $m$ is chosen as the largest possible value ($m = n$) for plots for $n$ up to 9 (top two rows), and $m = 10$ is chosen for larger $n$ values (bottom row). The "empty" region in the middle is clear for all $n$ greater than ~3.

We must allow that it could be possible that the object was very resonant, so then even some weak field penetrating to the object could still excite significant effective sources in the object, and that could somewhat increase the number of waves we had to consider in the scattering or absorption of that object. Even then, the fall-off of the penetration of higher *n* waves into the spherical bounding volume is quite rapid with *n*, so $N_p$ (or its good approximation $N_{SH}$) is still a good first guide to the number of waves that can couple strongly in or out of the object.

In conclusion, when we consider incoming spherical waves, we find that those also experience the "escape radius", which now becomes the start of a tunneling barrier that may prevent them from penetrating much further *into* the spherical volume. Such waves can effectively be largely or essentially entirely reflected by this spherical volume in free space, leading to standing waves that could appear to us to have simply "focused" through the volume. With increasing *n*, these waves cannot sense objects of size much smaller than the spherical surface of radius given by the escape radius $r_{escn}$. So, indeed, there are also only so many waves that can effectively get into a volume, e.g., for active sensing, and this counting is essentially the same as for the number that are well coupled out of the volume.

## S5 Circular and cylindrical waves

Though spherical waves give the most general results on bounding volumes and counting of waves from them, cylindrical and (planar) circular waves also occur in many contexts. Importantly, in such coordinates we also see similar wave tunneling and escape behavior in the radial parts of these coordinate systems, and for completeness we derive that here. We can proceed in a similar fashion in polar or cylindrical coordinates to that taken in spherical coordinates.

### S5.1 Scalar circular and cylindrical waves

We start with scalar waves with the same scalar Helmholtz equation (Eq. (24). We introduce cylindrical coordinates $r_{xy}$, $\phi$ and *z*. Considering the usual Cartesian coordinates *x*, *y*, and *z* as a reference coordinate system, *z* is now also the coordinate along the axis of cylindrical coordinates, $r_{xy}$ is the radial coordinate in any plane parallel to the *x-y* plane, and $\phi$ is the angular (azimuthal) coordinate in any *x-y* plane. (We have used $\phi$ rather than $\theta$ for this angle for commonality with the spherical coordinate system.) For circular (or polar) coordinates, corresponding to waves in planar or two-dimensional structures or systems, there is no variation in *z* and that axis and coordinate can be dropped.

In such coordinates, the scalar Helmholtz equation becomes

$$\nabla^2 U + k^2 U \equiv \left( \frac{1}{r_{xy}} \frac{\partial}{dr_{xy}} r_{xy} \frac{\partial}{\partial r_{xy}} + \frac{1}{r_{xy}^2} \frac{\partial^2}{\partial \phi^2} + \frac{\partial^2}{\partial z^2} \right) U + k^2 U = 0 \qquad (176)$$

For completeness in our notation, we formally give the algebraic derivation of the separation of this equation. We propose a separable solution of the form

$$U(\mathbf{r}) = R_{xy}(r_{xy})\Phi(\phi)Z(z) \tag{177}$$

Dividing by $-R_{xy}(r_{xy})\Phi(\phi)Z(z)$ and rearranging gives

$$\frac{-1}{R_{xy}(r_{xy})\Phi(\phi)}\left(\frac{1}{r_{xy}}\frac{\partial}{dr_{xy}}r_{xy}\frac{\partial}{\partial r_{xy}} + \frac{1}{r_{xy}^2}\frac{\partial^2}{\partial\phi^2}\right)R_{xy}(r_{xy})\Phi(\phi) = \frac{1}{Z}\frac{\partial^2 Z}{\partial z^2} + k^2 = k^2 - k_z^2 \tag{178}$$

where we have introduced the separation constant $k^2 - k_z^2$. The resulting equation for Z has solutions of the form

$$Z(z) = \exp(ik_z z) \tag{179}$$

Multiplying Eq. (178) by $r_{xy}^2$ and rearranging gives

$$\frac{-r_{xy}^2}{R_{xy}(r_{xy})}\left(\frac{1}{r_{xy}}\frac{\partial}{dr_{xy}}r_{xy}\frac{\partial}{\partial r_{xy}}\right)R_{xy}(r_{xy}) - (k^2 - k_z^2)r_{xy}^2 = \frac{1}{\Phi(\phi)}\frac{\partial^2}{\partial\phi^2}\Phi(\phi) = -n^2 \tag{180}$$

where we have introduced the separation constant $-n^2$. The resulting equation for $\Phi$ becomes

$$\frac{\partial^2}{\partial\phi^2}\Phi(\phi) = -n^2\Phi(\phi) \tag{181}$$

which has solutions of the form

$$\Phi(\phi) = \exp(in\phi) \text{ for } n = \ldots -2, -1, 0, 1, 2\ldots \tag{182}$$

i.e., for $n$ being a positive or negative integer or zero. This requirement on $n$ is so we obtain a single-valued function for $\Phi$. Finally, dividing Eq. (180) by $-r_{xy}^2$ and rearranging gives

$$\left(\frac{1}{r_{xy}}\frac{\partial}{dr_{xy}}r_{xy}\frac{\partial}{\partial r_{xy}}\right)R_{xy}(r_{xy}) + \left(k_{xy}^2 - \frac{n^2}{r_{xy}^2}\right)R_{xy}(r_{xy}) = 0 \tag{183}$$

where we have defined

$$k_{xy} = \sqrt{k^2 - k_z^2} \tag{184}$$

(taking the positive square root for definiteness). Changing to the variable

$$\rho_{xy} = k_{xy}r_{xy} \tag{185}$$

with $C_n(\rho_{xy}) \equiv R_{xy}(r_{xy})$, Eq. (183) becomes

$$\left(\frac{1}{\rho_{xy}}\frac{\partial}{d\rho_{xy}}\rho_{xy}\frac{\partial}{\partial\rho_{xy}}\right)C_n(\rho_{xy}) + \left(1 - \frac{n^2}{\rho_{xy}^2}\right)C_n(\rho_{xy}) = 0 \tag{186}$$

which is one way of writing the standard differential equation for Bessel functions. Specific solutions include the Bessel functions of the first and second kinds, $J_n(\rho_{xy})$ and $Y_n(\rho_{xy})$, respectively, and the linear combinations $H_n^{(1)}(\rho_{xy}) = J_n(\rho_{xy}) + iY_n(\rho_{xy})$ and $H_n^{(2)}(\rho_{xy}) = J_n(\rho_{xy}) - iY_n(\rho_{xy})$, which are the (cylindrical) Hankel functions of the first and second kinds, respectively, conventionally corresponding to outward and inward circular waves, respectively. Collectively, these four different solutions are sometimes known as the cylinder functions, hence the notation $C_n$ to represent any of them.

Now, we can define a new set of functions

$$K_n(s) = \sqrt{s}\, C_n(s) \tag{187}$$

The relation of these functions to the circular (or Bessel or Hankel) functions is conceptually similar to the relation of the Riccati-Bessel functions to spherical Bessel or Hankel functions. (We are not aware of any corresponding name for these functions $K_n(s)$, however.) In the present case, we are multiplying by the square root of a radial distance $s$ to take out an underlying $1/\sqrt{s}$ behavior of the Bessel or Hankel functions at large radius, a behavior that corresponds to Bessel or Hankel functions representing circularly expanding waves at large radius. If we substitute $K_n(\rho_{xy})/\sqrt{\rho_{xy}}$ for $C_n(\rho_{xy})$ in Eq. (186), we obtain

$$-\frac{\partial^2 K_n(\rho_{xy})}{\partial \rho_{xy}^2} + \frac{(n^2 - 1/4)}{\rho_{xy}^2} K_n(\rho_{xy}) = K_n(\rho_{xy}) \tag{188}$$

Just like for the spherical case with Riccati-Bessel functions, this is in the form of a Schrödinger equation with a "potential energy"

$$V_{xy}(\rho_{xy}) = \frac{(n^2 - 1/4)}{\rho_{xy}^2} \tag{189}$$

and an "eigen energy" of $E_n = 1$, similar to Eq. (77) for the spherical wave case. (Note also that this potential is proportional to 1/(radius)$^2$, just as in the spherical case.) The corresponding escape radius $\rho_{xyen}$ in dimensionless form is given by the condition $\rho_{xyen}^2 = n^2 - \frac{1}{4}$, i.e.,

$$\rho_{xyen} = \sqrt{n^2 - \tfrac{1}{4}} \tag{190}$$

In dimensioned form, using Eqs. (184) and (185), we have the escape radius

$$r_{xyen} = \frac{\sqrt{n^2 - \tfrac{1}{4}}}{\sqrt{k^2 - k_z^2}} \tag{191}$$

For the case of circular waves, such as those on a membrane, there is no variation in $z$. Then the $\partial^2/\partial z^2$ term in Eq. (176) and any $Z(z)$ function can be omitted, and the $k_z$ can be dropped from Eq. (191), giving a simpler circular escape radius

$$r_{cen} = \left(\sqrt{n^2 - \tfrac{1}{4}}\right)/k \tag{192}$$

Hence, these $K_n(\rho_{xy})$ cylindrical or circular waves have a similar set of tunneling, escape and propagating wave behaviors to the Riccati-Bessel spherical waves, with a corresponding escape radius. Note in cylindrical or circular cases that for $n = 0$, the waves all start out by propagating – there is then no finite escape radius.

Note that the escape radius in the cylindrical case depends on any $k_z$ value. Larger $k_z$ values will lead to larger escape radii. The smallest possible escape radius for any real $k_z$ is therefore given by the circular escape radius $r_{cen}$ of Eq. (192).

## S5.2 Vector circular and cylindrical waves

The general analysis of vector fields in section S2.2 above is independent of the coordinate system, so we can if we wish follow through with analysis in **L**, **M**, and **N** waves in cylindrical coordinates also (see, for example, Stratton [30], pp. 395-399). Formally, for whatever cylinder function $C_n$ we choose as appropriate for our problem (i.e., Bessel or Hankel functions), with $\hat{\mathbf{z}}$ as a unit vector in the $z$ direction and writing

$$\psi_n \equiv C_n(k_{xy} r_{xy}) \exp(in\phi) \exp(ik_z z) \tag{193}$$

we can write (e.g., as in [33], p. 23)

$$\mathbf{L}_n(k_{xy}, k_z, \mathbf{r}) = \nabla \psi_n \tag{194}$$

$$\mathbf{M}_n(k_{xy}, k_z, \mathbf{r}) = \nabla \times (\hat{\mathbf{z}} \psi_n) \equiv \nabla \psi_n \times \hat{\mathbf{z}} \tag{195}$$

$$\mathbf{N}_n(k_{xy}, k_z, \mathbf{r}) = \frac{1}{k} \nabla \times \nabla \times (\hat{\mathbf{z}} \psi_n) \equiv \frac{1}{k} \nabla \times \mathbf{M}_n(k_{xy}, k_z, \mathbf{r}) \tag{196}$$

For electromagnetic waves there will still be no **L** waves as before. For each of the electric and magnetic fields, we will have solutions of the forms given by these $\mathbf{M}_n$ and $\mathbf{N}_n$ functions. In component form, with $\hat{\mathbf{r}}_{xy}$ as the radial unit vector in the direction of interest in the $x$-$y$ plane, from Eq. (195)

$$\mathbf{M}_n = \frac{in}{r_{xy}} \psi_n \hat{\mathbf{r}}_{xy} - \frac{\partial \psi_n}{\partial r_{xy}} \hat{\phi} \equiv \left[ \frac{in}{\rho_{xy}} C_n(\rho_{xy}) \hat{\mathbf{r}}_{xy} - \frac{\partial C_n(\rho_{xy})}{\partial \rho_{xy}} \hat{\phi} \right] k_{xy} \exp(in\phi + ik_z z) \tag{197}$$

and from Eq. (196)

$$\begin{aligned}\mathbf{N}_n &= \frac{1}{k} \left[ ik_z \frac{\partial \psi_n}{\partial r_{xy}} \hat{\mathbf{r}}_{xy} - \frac{nk_z}{r_{xy}} \psi_n \hat{\phi} + k_{xy}^2 \psi_n \hat{\mathbf{z}} \right] \\ &\equiv \left[ ik_z k_{xy} \frac{\partial C_n(\rho_{xy})}{\partial \rho_{xy}} \hat{\mathbf{r}}_{xy} - \frac{nk_z k_{xy}}{\rho_{xy}} C_n(\rho_{xy}) \hat{\phi} + k_{xy}^2 C_n(\rho_{xy}) \hat{\mathbf{z}} \right] \frac{1}{k} \exp(in\phi + ik_z z)\end{aligned} \tag{198}$$

If we take a similar approach to defining corresponding **E** and **H** fields as for the spherical case above, then we can formally choose

$$\mathbf{H}_n^{(TM)} = \mathbf{M}_n \tag{199}$$

which is polarized entirely in the *x-y* plane. (Formally, similarly to the spherical case, we would need to multiply by some constant with dimensions of amperes to give this the correct dimensions for this magnetic field.) In the far field, this will tend to being polarized in the $\hat{\phi}$ direction, transverse to the radial direction (the component of $\mathbf{M}_n$ in the $\hat{\mathbf{r}}_{xy}$ direction will fall off as $\sim 1/r_{xy}^{3/2}$ at large $r_{xy}$ and hence will become negligible, leaving just a $\hat{\phi}$ component). The corresponding electric field for this **H** field is

$$\mathbf{E}_n^{(TM)} = i\sqrt{\frac{\mu}{\varepsilon}} \mathbf{N}_n \tag{200}$$

Now if we choose

$$\mathbf{E}_n^{(TE)} = i\sqrt{\frac{\mu}{\varepsilon}} \mathbf{M}_n \tag{201}$$

which also is polarized entirely in the *x-y* plane and will tend to being polarized in the $\hat{\phi}$ direction in the far field, the corresponding **H** field is

$$\mathbf{H}_n^{(TE)} = \mathbf{N}_n \tag{202}$$

In these fields, again the radial behavior of the amplitude is governed by the $C_n(\rho_{xy})$ function, or, equivalently, the $K_n(\rho_{xy}) = \sqrt{\rho_{xy}}\, C_n(\rho_{xy})$ function that removes the underlying $1/\sqrt{\rho_{xy}}$ behavior of the circular functions at large radius, and which shows the same tunneling behavior as for the scalar circular or cylindrical case.

The actual behavior of these functions for finite $k_z$ is somewhat complicated. However, the situation for $k_z = 0$, which corresponds to the circular case also, becomes much simpler. In this case, the expression for $\mathbf{M}_n$ becomes

$$\mathbf{M}_{n,k_{xy}=0} = \left[ \frac{in}{\rho_{xy}} C_n(\rho_{xy}) \hat{\mathbf{r}}_{xy} - \frac{\partial C_n(\rho_{xy})}{\partial \rho_{xy}} \hat{\phi} \right] k \exp(in\phi) \tag{203}$$

and for $\mathbf{N}_n$ we have

$$\mathbf{N}_{n,k_z=0} = k\psi_n \hat{\mathbf{z}} \equiv k\, C_n(\rho_{xy}) \exp(in\phi) \hat{\mathbf{z}} \tag{204}$$

which is a field polarized along the *z* direction, and shows simple Bessel or Hankel function behavior with radius. The corresponding $K_n(\rho_{xy}) = \sqrt{\rho_{xy}}\, C_n(\rho_{xy})$ function will then directly show the tunneling and escape behavior.

## S5.3 Summary and comparison with the spherical case

We can summarize the following similarities and differences for the circular/cylindrical case compared to the spherical case.

1) In both cases, we can define functions for the radial behavior that satisfy Schrödinger-like equations, with a clearly defined escape radius. For the circular/cylindrical cases these functions are of the form $K_n(\rho_{xy}) = \sqrt{\rho_{xy}}\, C_n(\rho_{xy})$ (Eq. (187)), where $\rho_{xy}$ is the appropriate dimensionless radial variable (Eqs. (184) and (185)). Multiplying by the factor $\sqrt{\rho_{xy}}$ "takes out" the underlying $1/\sqrt{\text{radius}}$ behavior of a cylindrically or circularly expanding wave. Equivalently, the $K_n(\rho_{xy})$ can be regarded as waves normalized per unit (circular) angle rather than per unit (transverse) length.

2) The expression for the escape radius for a circular/cylindrical wave of order $n$, $\rho_{xyen} = \sqrt{n^2 - 1/4}$ (Eq. (190), is similar to that for the spherical case in its behavior with $n$. For circular waves, the dimensioned escape radius $r_{cen} = \left(\sqrt{n^2 - 1/4}\right)/k$ (Eq. (192)) is scaled using the overall wavevector magnitude $k$, similarly to the spherical case. For moderate or larger $n$, if we consider an effective wavelength of $2\pi/k_{xy}$ in the cylindrical case, we can interpret this as saying that, if the number of (effective) wavelengths round the circumference of the circle of interest exceeds $\sim n$, that wave must tunnel radially to escape.

3) For cylindrical waves with finite (real) wavevector $k_z$ along the $z$ direction, the radial wavevector component $k_{xy}$ has to reduce according to $k_{xy} = \sqrt{k^2 - k_z^2}$ (Eq. (184)) so that $k_{xy}^2 + k_z^2 = k^2$. As a result, the dimensioned escape radius for a given $n$ increases as $k_z$ increases, according to $r_{xyen} = \left(\sqrt{n^2 - 1/4}\right)/\left(\sqrt{k^2 - k_z^2}\right)$ (Eq. (191)). Hence,

   (i) $r_{cen} = \left(\sqrt{n^2 - 1/4}\right)/k$ is the smallest escape radius for a given $n$, and

   (ii) radial escape becomes more difficult as $k_z$ increases.

4) Unlike the spherical case, waves with $n = 0$ are allowed for both scalar and electromagnetic waves for both circular and cylindrical waves. As for the spherical case, $n = 0$ waves start out by propagating radially, with no escape radius.

So, in conclusion, circular and cylindrical waves show a similar kind of propagating and tunneling behavior in the radial direction to that for the spherical case, also with a clearly defined escape radius for a given order $n$ of wave.